\pdfoutput=1

\documentclass[12pt]{article}
\usepackage{jheppub}

\usepackage{amsmath}
\usepackage{amssymb}
\allowdisplaybreaks

\newcommand{\cG}{\mathcal{G}}
\newcommand{\cV}{\mathcal{V}}
\newcommand{\dd}{\mathrm{d}}
\newcommand{\cL}{\mathcal{L}}
\DeclareMathOperator{\Tr}{Tr}

\title{Three-dimensional vacua of $G_2$ flux compactifications} 

\author{Aravind Aikot$^a$, Julian Choi$^b$, and Timm Wrase$^a$}

\affiliation{$^a$Department of Physics, Lehigh University,\\
16 Memorial Drive East, Bethlehem, PA 18015, USA}

\affiliation{$^b$Department of Physics, Cornell University,\\
Ithaca, NY 14853, USA}

\emailAdd{ara425@lehigh.edu}
\emailAdd{yc2698@cornell.edu}
\emailAdd{timm.wrase@lehigh.edu}

\abstract{We study supersymmetric vacua of three-dimensional $\mathcal N=1$ effective theories from type IIA, type IIB/type I, and heterotic compactifications on $G_2$-structure spaces. Superpotential homogeneity and the physical scalar metric yield universal vacuum relations and axion stabilization constraints without specifying the full moduli-space metric. We express AdS$_3$ scalar masses through $\Delta$, the scaling dimension of the corresponding operator in the dual two-dimensional conformal field theory. In massive type IIA AdS$_3$ vacua, we derive the bound $\Delta_{\max}\geq(20+4\sqrt{67})/7\approx 7.5$ and an exact dimension $\Delta_b=6$ for $B_2$ axions on the torsion-free branch. Type IIB/type I and heterotic AdS$_3$ vacua contain a universal volume--dilaton mode with $\Delta_{\phi s}=4$. On the singlet-torsion type IIB branch, we also obtain the volume--dilaton mode with $\Delta_{\phi s}=20/7$ and $\Delta_b=16/7$. Every supersymmetric Minkowski vacuum in the retained classical closed-string theories possesses an exact volume--dilaton flat direction. We characterize physical large-flux limits after tadpole constraints and axionic identifications, and identify torsion obstructions to isotropic scale separation.}

\begin{document}

\maketitle

\section{Introduction}
\label{sec:introduction}

Flux compactifications provide a direct mechanism for generating scalar potentials and stabilizing moduli in lower-dimensional string theory. In three dimensions, they also furnish a comparatively less-explored setting for the AdS/CFT duality and the question of whether the AdS and Kaluza--Klein scales can be parametrically separated. Compactifications on seven-manifolds with $G_2$ holonomy, or more generally with $G_2$ structure, are natural for this purpose because they preserve minimal supersymmetry and permit flux, intrinsic torsion, and localized sources to enter a real superpotential~\mbox{\cite{Farakos:2020phe, Emelin:2021gzx, Coudarchet:2023mfs}}.

General scaling arguments provide guidance for this investigation. When the internal curvature contributes at leading order to the vacuum energy, it generically ties the Kaluza--Klein and lower-dimensional curvature scales together in an isotropic compactification, preventing a parametrically separated low-energy theory. By contrast, a leading O$p$-plane contribution with $p<7$ drives scale separation in controlled weak-coupling or large-volume limits~\cite{Gautason:2015tig,Tringas:2025uyg}. These results make it important to distinguish leading from subleading curvature and source terms, and isotropic from anisotropic limits.

Although our focus is on ten-dimensional string theory, closely related M-theory compactifications on $G_2$ spaces established the flux superpotential and moduli-fixing framework~\cite{Gukov:1999ya, Beasley:2002db, Acharya:2002kv}. Supersymmetry conditions and their relation to the intrinsic torsion were developed in~\cite{BehrndtJeschekGstructures2003, KasteMinasianTomasiello2003, BehrndtJeschekSuperpotential2004,House:2004pm}, while geometric-flux reductions and vacuum classifications were studied in~\cite{DallAgata:2005zlf, BehrndtCveticLiu2006, McOrist:2012yc}. A recent weak-$G_2$ analysis gives a complementary perspective on scale separation~\cite{VanHemelryck:2024bas}. A real superpotential for warped M-theory compactifications on eight-manifolds with local $G_2$ structure was recently derived without a finite Kaluza--Klein truncation in~\cite{Ashmore:2026vgf}.

The first systematic massive-IIA constructions on $G_2$ orientifolds exhibited no-scale Minkowski vacua and controlled AdS$_3$ families~\cite{Farakos:2020phe}. Subsequent work developed complementary supersymmetry formalisms, additional flux sectors, and detailed scale-separated examples~\cite{VanHemelryck:2022ynr,Farakos:2023nms,Farakos:2023wps,Farakos:2025bwf}. Type IIB orientifolds and broader three-dimensional classifications revealed both supersymmetric and non-supersymmetric families, as well as the importance of open-string fields in assessing stability~\cite{Emelin:2021gzx,Arboleya:2024vnp,VanHemelryck:2025qok,Arboleya:2025tax,Arboleya:2025lwu}. A related duality program connects massive-IIA solutions to type I and heterotic frames and has produced controlled anisotropic heterotic examples, massless-IIA families, and solutions with extended supersymmetry~\cite{Miao:2025rgf,Tringas:2025bwe,Tringas:2026ncg,Cribiori:2026caf}. Related constructions on $G_2$-structure spaces times a circle yield scale-separated AdS$_2$ flux vacua in type II string theory~\cite{Tringas:2026ptt}.

Most explicit vacuum analyses nevertheless use highly symmetric orbifold truncations. In particular, sectors associated with nontrivial two- and five-forms are often absent, even though they contain axions and three-dimensional vectors whose St\"uckelberg couplings change the physical moduli count. The complete effective actions required to retain these sectors, including the axion-dressed field strengths, geometric fluxes, kinetic mixing, and type I and heterotic gauge bundles, were derived in~\cite{Aikot:2026tzn}. The purpose of the present paper is to use those actions to study their vacua and spectra without assuming an explicit metric on the entire $G_2$ moduli space.

Our analysis addresses three questions: which vacuum relations follow from dilaton and volume homogeneity alone; what conditions on the geometry and background fluxes are required for moduli stabilization; and which scalar masses, and hence conformal dimensions of the putative dual operators, are fixed independently of the details of the compactification. Treating the three string-theory frames in one framework permits a direct comparison of flux, intrinsic torsion, and localized sources.

Section~\ref{sec:general-vacua} establishes the common supergravity, $G_2$, axion, and mass conventions. Sections~\ref{sec:type-IIA}, \ref{sec:type-IIB-type-I}, and~\ref{sec:heterotic} analyze the type IIA, type IIB/type I, and heterotic theories, respectively. We collect several of our results and discuss future directions in section \ref{sec:conclusions}.

\section{\texorpdfstring{Three-dimensional $\mathcal{N}=1$ supergravity and $G_2$ reductions}{Three-dimensional N=1 supergravity and G2 reductions}}
\label{sec:general-vacua}

We first collect the geometric and three-dimensional supergravity results used below. We work at two-derivative order in the universal bulk sectors, retaining the leading $\alpha'$ gauge, curvature, and Green--Schwarz terms needed for the heterotic and type I analyses. All reductions are finite dimensional, and vacuum and mass conditions are imposed on gauge-inequivalent fields unless stated otherwise.

\subsection{\texorpdfstring{$G_2$ notation}{G2 notation}}
\label{subsec:g2-notation}

Let $X$ be an oriented smooth seven-manifold with a $G_2$ structure. The associative three-form $\Phi$ determines a metric $g(\Phi)$, a Hodge operator $\star_{g(\Phi)}$, and the coassociative four-form
\begin{equation}
    \Psi\equiv\star\Phi\,.
\end{equation}
The internal volume is then defined
\begin{equation}
    \cV\equiv\frac{1}{7}\int_X\Phi\wedge\Psi\,.
\end{equation}
Unless a string-frame subscript $s$ is displayed, the internal metric, Hodge operator, $G_2$ forms, and $\cV$ are understood as ten-dimensional Einstein-frame quantities. 

The failure of $\Phi$ and $\Psi$ to be closed can be written in terms of four torsion classes~\cite{FernandezGray,2003math......5124B},
\begin{align} \label{taus}
    \dd_7\Phi&=\tau_0\Psi+3\tau_1\wedge\Phi+\star\tau_3\,,\\
    \dd_7\Psi&=4\tau_1\wedge\Psi+\tau_2\wedge\Phi\,.
\end{align}
Here, $\tau_0\in\Omega^0_1(X),\tau_1\in\Omega^1_7(X),\tau_2\in\Omega^2_{14}(X),\tau_3\in\Omega^3_{27}(X)$, where $\Omega^p_r(X)$ is the subspace of $p$-forms in the $r$-dimensional irreducible representation of $G_2$. In this paper, we consider compact $G_2$-structure spaces that have no retained 1-forms (thus, $\tau_1=0$).

For the compactifications studied in this paper, our $G_2$-structure manifold has generically nonvanishing torsion and a consistent truncation to a finite-dimensional subspace of forms. We assume this subspace is mapped to itself under $\dd_7$. We choose bases of 2-, 3-, 4- and 5-forms and denote them $\{\Upsilon_A\}$, $\{\Phi_I\}$, $\{\Psi^I\}$, and $\{T^A\}$, respectively, where $A,I$ index the number of 2- and 3-forms. The bases are normalized such that
\begin{equation}
    \int_X\Upsilon_A\wedge T^B=\delta_A^B\,,
    \qquad
    \int_X\Phi_I\wedge\Psi^J=\delta_I^J\,.
    \label{orthonormality}
\end{equation}
The bases form 2-, 3-, and 4-form metrics, which we define as
\begin{equation}
    G_{AB}\equiv\frac 1{\mathcal V}\int_X\Upsilon_A\wedge\star\Upsilon_B\,,\qquad G_{IJ}\equiv\frac 1{\mathcal V}\int_X\Phi_I\wedge\star\Phi_J\,,\qquad \widehat{G}^{IJ}\equiv\frac 1{\mathcal V}\int_X\Psi^I\wedge\star\Psi^J\label{moduli-metric}\,.
\end{equation}
We also define triple-intersection numbers as follows.
\begin{equation}
    \kappa_{ABI}\equiv\int_X\Upsilon_A\wedge\Upsilon_B\wedge\Phi_I.\label{triple-intersection}
\end{equation}
Note that the triple-intersection numbers are purely topological quantities.

Lastly, we define the non-closure matrices by
\begin{equation}
    \dd_7\Upsilon_A\equiv N_A{}^I\Phi_I\,,
    \qquad
    \dd_7\Phi_I\equiv M_{IJ}\Psi^J\,,
    \qquad
    \dd_7\Psi^J\equiv Q^J{}_A T^A\,.
    \label{torsion-matrices} 
\end{equation}
Nilpotency and integration by parts, assuming compactness, imply
\begin{equation}
    N_A{}^I M_{IJ}=0\,,
    \qquad
    M_{IJ}Q^J{}_A=0\,,
    \qquad
    Q^J{}_A=-N_A{}^J\,,
    \qquad
    M_{IJ}=M_{JI}\,.
    \label{nilpotency}
\end{equation}
We similarly obtain from $\int_X \dd_7(\Upsilon_A \wedge \Upsilon_B \wedge \Upsilon_C) = 0$ that 
\begin{equation}
    N_A{}^I\kappa_{BCI} + N_B{}^I\kappa_{ACI} + N_C{}^I\kappa_{ABI} = 0\,.\label{stokes}
\end{equation}

In a harmonic truncation on a manifold with full $G_2$ holonomy, the form bases are necessarily closed and their indices range $A=1,\ldots,b^2(X)=b^5(X)$ and $I=1,\ldots,b^3(X)=b^4(X)$, where $b^r(X)$ is the $r$-th Betti number of $X$. On such a compact manifold with $G_2$ holonomy, harmonic two-forms lie in $\Lambda^2_{14}$. They therefore satisfy the algebraic $G_2$ identity $\star\Upsilon_A=-\Upsilon_A\wedge\Phi$~\cite{JoyceBookSpecialHolonomy,KarigiannisDeformations}. Using the expansion $\Phi=s^I\Phi_I$ together with \eqref{moduli-metric} and \eqref{triple-intersection}, we obtain
\begin{equation}
G_{AB}=-\frac{1}{\mathcal V}\kappa_{ABI}s^I .
\label{eq:metric-intersection-identity}
\end{equation}
The same derivation applies on a general $G_2$-structure space whenever the retained two-forms lie pointwise in $\Lambda^2_{14}$. Intrinsic torsion does not by itself modify this algebraic identity, and below we assume that the retained two-form basis satisfies this representation-theoretic condition, as in \cite{Aikot:2026tzn}.

\subsection{AdS and Minkowski vacua}
\label{subsec:ads-minkowski}

The bosonic part of three-dimensional $\mathcal{N}=1$ supergravity coupled to real scalar multiplets can be written as~\cite{Emelin:2021gzx}
\begin{equation}
    S_3
    =
    4\pi\int\dd^3x\,\sqrt{-g_3}
    \left[
        \frac{1}{2}R_3
        -\cG_{MN}(\varphi)
        \partial_\mu\varphi^M\partial^\mu\varphi^N
        -V(\varphi)
    \right]
    +\ldots\,,
    \label{eq:general-3d-action}
\end{equation}
where $\cG_{MN}$ is positive definite on the physical scalar manifold and $\varphi^M$ denotes all real scalar fields: the dilaton $\phi$, saxions $s^I$, $B_2$/RR axions, and scalars dual to three-dimensional vectors. The collective scalar field metric $\mathcal G_{MN}(\varphi)$ is schematically block-diagonal with the following universal sectors~\cite{Aikot:2026tzn}:
\begin{equation}
    \mathcal G_{\phi\phi}=\frac 14\,,\qquad \mathcal G_{IJ}=\frac 14G_{IJ}+\frac 1{36\mathcal V^2}\left(\int_X\Phi_I\wedge\Psi\right)\left(\int_X\Phi_J\wedge\Psi\right)\,.\label{scalar-field-metric}
\end{equation}

The scalar potential is determined by a real superpotential $P(\varphi)$,
\begin{equation}
    V
    =
    \cG^{MN}\partial_M P\,\partial_N P-4P^2\,.
    \label{eq:general-real-superpotential-potential}
\end{equation}
For configurations with vanishing three-dimensional gauge-field strengths, a supersymmetric vacuum satisfies
\begin{equation}
    \left.\partial_M P\right|_{\varphi=\varphi_\star}=0
    \label{eq:general-supersymmetry-condition}
\end{equation}
for every gauge-inequivalent scalar direction. In a formulation containing St\"uckelberg fields, the superpotential is gauge invariant and \eqref{eq:general-supersymmetry-condition} is imposed after removing the gauged directions.

Using the Levi--Civita connection of $\cG_{MN}$, one finds
\begin{equation}
    \nabla_M V
    =
    2\left(\nabla_M\partial_N P\right)
    \cG^{NK}\partial_K P
    -8P\partial_M P\,,
    \label{eq:general-gradient-potential}
\end{equation}
which vanishes at any solution of \eqref{eq:general-supersymmetry-condition}. At such a point,
\begin{equation}
    V_\star=-4P_\star^2\leq 0\,.
    \label{eq:general-susy-vacuum-energy}
\end{equation}
Consequently, a supersymmetric solution with $P_\star\neq0$ is AdS$_3$, with
\begin{equation}
    V_\star=-\frac{1}{L_{\mathrm{AdS}}^2}\,,
    \qquad
    L_{\mathrm{AdS}}^{-2}=4P_\star^2\,.
    \label{eq:general-AdS-radius}
\end{equation}
A supersymmetric Minkowski vacuum instead requires
\begin{equation}
    P_\star=0\,,
    \qquad
    \left.\partial_M P\right|_\star=0\,.
    \label{eq:general-susy-Minkowski}
\end{equation}
Supersymmetric de Sitter vacua are excluded in the two-derivative theory. This does not exclude non-supersymmetric de Sitter critical points, which would have to satisfy $\partial_MV=0$ and $V_\star>0$ and whose stability must be checked using the complete physical mass matrix.

\subsection{Physical axions and equivalent flux vectors}
\label{subsec:physical-axions}

The number of coefficients appearing in an expansion of the ten-dimensional potentials
is not the number of physical axionic moduli by itself. Geometric flux can gauge
continuous axionic shift symmetries; when an axion $c^I$ and a vector $A^A$ arise from
the same higher-form potential, the non-closure relation in \eqref{torsion-matrices} implies that the naive field strength built from $c^I$ alone is not gauge invariant, and
must be completed to
\begin{equation}
    Dc^I\equiv\dd c^I-N_A{}^I A^A\,.
    \label{eq:general-covariant-derivative}
\end{equation}
This covariant derivative defines a genuine $\mathrm U(1)^{\dim A}$ gauge symmetry, under
which $A^A$ transforms in the usual way and $c^I$ shifts by a compensating amount fixed
by the same matrix $N_A{}^I$:
\begin{equation}
    \delta A^A=\dd\lambda^A\,,
    \qquad
    \delta c^I=N_A{}^I\lambda^A\,,
    \label{eq:general-Stueckelberg-coupling}
\end{equation}
with $\lambda^A(x)$ the local gauge parameter. 

Thus, any combination $\delta c^I\in\operatorname{im}N_A{}^I$ is eaten by $A^A$, which becomes massive St\"uckelberg fields. The vector combinations in $\ker N$ remain ungauged and can be dualized to scalars in three dimensions, given the absence of obstructing Chern--Simons couplings. In type IIB, the vectors from $C_4$ are already dual to the axions $c_4^\rho$ by self-duality of $\widetilde F_5$, so dualizing them adds no scalars.

It is also necessary to distinguish eaten axions from physical axions that are absent from the perturbative superpotential. Suppose, as in the type IIA sector below, that the only dependence on $c^I$ occurs through $c^I M_{IJ}$. A shift $\delta c^I$ is then perturbatively flat when $\delta c^I M_{IJ}=0$. The first relation in \eqref{nilpotency} guarantees that every gauged direction lies in this kernel. The perturbatively flat physical axions in this sector therefore take values in
\begin{equation}
    H^3_\text{tr}(X)=\ker M_{IJ}\,/\,\text{im }N_A{}^I\,,
    \label{eq:general-physical-flat-axions}
\end{equation}
the cohomology of the finite-dimensional complex of retained forms at $\Omega^3_{\mathrm{tr}}(X)$. Similarly, two-form axions that enter only through $\dd(\delta b^A\Upsilon_A)$ are flat along $\ker N_A{}^I$, while four-form axions that enter only through $\dd(\delta c_{4,I}\Psi^I)$ are flat along $\ker Q^J{}_A$. The precise parity restrictions and possible additional flux couplings will be specified separately for type IIA, type IIB/type I, and heterotic theory.

Because the complete off-shell superpotential is independent of these gauge inequivalent shifts $\delta c\in H^3_\text{tr}(X)$, a supersymmetric solution extends to a continuous family along them at two-derivative order. The corresponding axions are therefore flat to all orders in the perturbative axion dependence of the retained superpotential, rather than only at quadratic order. Higher-derivative or nonperturbative effects are expected to lift them, whereas an eaten axion is not a scalar modulus in the low-energy spectrum. We will therefore distinguish throughout between stabilized axions, perturbatively flat axions, and St\"uckelberg fields.

The same truncated cohomology also determines whether a flux parameter is physically independent. Suppose that an internal gauge-invariant field strength contains the axion-dressed coefficient $f_J+c^I M_{IJ}$. Its decomposition into a background flux representative and an axion-dependent exact contribution is unchanged under
\begin{equation}
    f_J
    \longrightarrow
    f_J+k^I M_{IJ}\,,
    \qquad
    c^I
    \longrightarrow
    c^I-k^I\,.
    \label{eq:general-flux-axion-representative-shift}
\end{equation}
Thus, flux vectors that differ by an element of $\operatorname{im}M_{IJ}$ are physically equivalent, since their difference can be absorbed into the axions. If the truncation has a compatible integral flux lattice, the $k^I$ in \eqref{eq:general-flux-axion-representative-shift} are integer large shifts. An unbounded large-flux family must therefore come from a non-torsion integral class that survives the Bianchi identities and tadpole constraints, while exact classes can be removed by shifting the axions. 

\subsection{Breitenlohner--Freedman bound and conformal dimensions}
\label{subsec:masses-and-dimensions}

At a supersymmetric vacuum, the physical scalar mass operator is, per \eqref{eq:general-3d-action},
\begin{equation}
    \left(\mathcal{M}^2\right)^M{}_N
    =
    \left.
    \frac{1}{2}\cG^{MP}\nabla_P\partial_N V
    \right|_\star\,.
    \label{eq:general-physical-mass-operator}
\end{equation}
At a supersymmetric critical point, the covariant Hessian of the potential is
\begin{equation}
    \left.\nabla_M\partial_N V\right|_\star
    =
    2\left(\nabla_M\partial_P P\right)_\star
    \cG^{PQ}_\star
    \left(\nabla_Q\partial_N P\right)_\star
    -8P_\star
    \left(\nabla_M\partial_N P\right)_\star\,.
    \label{eq:general-potential-Hessian}
\end{equation}
It is convenient to regard the superpotential Hessian
\begin{equation}
    \mathcal{H}^M{}_N
    =
    \left.
    \cG^{MP}\nabla_P\partial_N P
    \right|_\star
    \label{eq:general-superpotential-Hessian}
\end{equation}
as a self-adjoint operator with respect to $\cG_{MN}$. Equations \eqref{eq:general-physical-mass-operator} and \eqref{eq:general-potential-Hessian} then give
\begin{equation}
    \mathcal{M}^2
    =
    \mathcal{H}^2-4P_\star\mathcal{H}\,.
    \label{eq:general-mass-from-superpotential-Hessian}
\end{equation}
This relation and the resulting AdS$_3$ Breitenlohner--Freedman bound were also discussed in~\cite{Ashmore:2026vgf}. If $\lambda$ is an eigenvalue of $\mathcal{H}$ at a supersymmetric AdS$_3$ vacuum, we find using \eqref{eq:general-AdS-radius} that the corresponding scalar mass is 
\begin{equation}
    m^2L_{\mathrm{AdS}}^2
    =
    \left(\frac{\lambda}{2P_\star}-1\right)^2-1\,.
    \label{eq:general-normalized-Hessian-mass}
\end{equation}
This form makes the AdS$_3$ Breitenlohner--Freedman bound~\cite{Breitenlohner:1982bm,Breitenlohner:1982jf} manifest. Thus, the scalar sector of a supersymmetric AdS$_3$ vacuum is perturbatively stable, even when the potential has negative Hessian eigenvalues.

For the standard AdS$_3$ quantization, the conformal dimension of the dual scalar operator satisfies
\begin{equation}
    \Delta(\Delta-2)=m^2L_{\mathrm{AdS}}^2\,,
    \qquad
    \Delta
    =
    1+\sqrt{1+m^2L_{\mathrm{AdS}}^2}
    =
    1+\left|\frac{\lambda}{2P_\star}-1\right|\,.
    \label{eq:general-conformal-dimension}
\end{equation}
Exact integer conformal dimensions are therefore equivalent to integer eigenvalues of the normalized superpotential Hessian; notably, a perturbatively flat axion has $\lambda=0$ and $\Delta=2$. This provides a direct diagnostic for the integer-dimension patterns observed in the DGKT vacua~\cite{Conlon:2021cjk,Apers:2022tfm} and in three-dimensional flux compactifications~\cite{Farakos:2025bwf,Arboleya:2025ocb}. Such integer spectra are not generic: the earlier massive-IIA AdS$_3$ examples analyzed in~\cite{Apers:2022zjx} have noninteger saxion dimensions.

\subsection{Kaluza--Klein scales and scale separation}
\label{subsec:general-control}

Let $m_{\mathrm{KK}}$ denote the lightest nonzero mass among the Kaluza--Klein towers after conversion to the three-dimensional Einstein frame of \eqref{eq:general-3d-action}, and $L_{\mathrm{KK}}=m_{\mathrm{KK}}^{-1}$. Since $L_{\mathrm{AdS}}^{-2}=|V_\star|$, scale separation at an AdS$_3$ vacuum requires, in addition to $g_s=e^\phi\ll1$ and internal radii large in string units,
\begin{equation}
    \frac{L_{\mathrm{KK}}^2}{L_{\mathrm{AdS}}^2}
    =
    \frac{|V_\star|}{m_{\mathrm{KK}}^2}
    \longrightarrow0\,.
    \label{eq:general-scale-separation}
\end{equation}
For $\dd s_{10}^2=e^{2A}\dd s_3^2+g_{mn}\dd y^m\dd y^n$ the Einstein-frame condition is $e^A\cV=1$, so an eigenvalue $\Lambda$ of the internal Laplacian appears in three dimensions as $m^2=\cV^{-2}\Lambda$. For an isotropic internal metric $\Lambda\sim\cV^{-2/7}$, and therefore
\begin{equation}
    m_{\mathrm{KK}}^2\sim\cV^{-16/7}\,.
    \label{eq:general-KK-scaling}
\end{equation}
For an anisotropic metric, the lightest mode is instead set by the largest radius, $\Lambda\sim R_{\max}^{-2}$, rather than by $\cV^{1/7}$ or the geometric mean of the radii, so $\cV\gg1$ alone does not guarantee \eqref{eq:general-scale-separation} and every radius must be tracked~\cite{Coudarchet:2023mfs}. A leading internal-curvature term in $V_\star$ scales as $\cV^{-16/7}$ in the isotropic case, which fixes \eqref{eq:general-scale-separation} to an order-one number~\cite{Gautason:2015tig,Tringas:2025uyg}. This is avoided in the massive-IIA large-$F_4$ family of section~\ref{subsec:iia-large-f4} and realized in the isotropic type IIB, type I, and heterotic cases of section~\ref{subsec:iib-flux-balance}.

\section{Type IIA}
\label{sec:type-IIA}

In this section, we study supersymmetric vacua of the three-dimensional low-energy effective theory for type IIA flux compactifications on $G_2$-structure spaces with O2/O6-planes, following~\cite{Aikot:2026tzn}. Although the full $G_2$ moduli-space metric is not known, we can nonetheless derive a universal relation among the flux contributions to the scalar potential and analyze the perturbative axion spectrum, the physical large-$F_4$ flux lattice and its parametric limits, and show that all supersymmetric Minkowski vacua necessarily have an exact classical flat direction.

\subsection{Tadpole conditions and equations of motion}
\label{subsec:iia-setup}

The O2/O6 orientifold projection and its action on the $G_2$ forms are given by~\cite{Angelantonj:2002ct,Grimm:2004ua,Aikot:2026tzn}
\begin{equation}\label{eq:O2O6projection}
    \mathcal{O}
    \equiv
    \Omega_p(-1)^{F_L}\sigma\,,
    \qquad
    \sigma^*\Phi=-\Phi\,,
    \qquad
    \sigma^*\Psi=+\Psi\,,
\end{equation}
where $\sigma:X\to X$ is the involutive isometry of the target space, $\Omega_p$ is the worldsheet parity, and $F_L$ is the left-moving fermion number operator. The orientifold projection splits our basis 2-, 3-, 4-, 5-forms into even and odd subspaces
\begin{equation}
    \begin{aligned}
    \Upsilon_A &\to \{\Upsilon_\alpha^+, \Upsilon_a^-\}\,,  &\Phi_I &\to \{\Phi^+_i, \Phi^-_\rho\}\,,\\
    \Psi^I &\to \{\Psi_+^\rho, \Psi_-^i\}\,, &T^A &\to \{T^{a}_+, T^{\alpha}_-\}\,,
    \end{aligned}\label{Projected-basis}
\end{equation}
where $\alpha=1,\cdots,b^2_+$, $a=1,\cdots,b^2_-$, $i=1,\cdots,b^3_+$, and $\rho=1,\cdots,b^3_-$. The 4- and 5-forms have been defined consistently with the oddness of the internal volume 7-form. The even- and odd-orientifold subspaces of the torsion matrices, the triple-intersection numbers, and the metric are denoted with Greek and Roman indices. 

After restricting to the appropriate parity sectors,~\eqref{orthonormality}--\eqref{eq:metric-intersection-identity} apply to the orientifolded space. 

We expand all the fields and fluxes in the appropriately transforming bases. The surviving $G_2$ structure is expanded in odd three-forms, while its Hodge dual is expanded in even four-forms:
\begin{equation} \label{eq:G2expansionO2O6}
    \Phi=s^\rho\Phi^-_\rho,\qquad\Psi=\psi_\rho(s)\Psi^\rho_+,\qquad \psi_\rho(s)=3\partial_\rho\cV.
\end{equation}
The last equality follows from the first variation of the Hitchin volume functional, with our normalization~\cite{HitchinStableForms}. The surviving components of the ten-dimensional potentials are expanded as
\begin{equation}
    B_2=b^a\Upsilon_a^-\,,
    \qquad
    C_3=c^i\Phi_i^+
    +A^\alpha\wedge\Upsilon_\alpha^+\,.
    \label{eq:iia-potential-expansions}
\end{equation}
The quantized background fluxes are expanded as 
\begin{equation}
    H_3^{\mathrm{bg}}=h^\rho\Phi_\rho^-\,,
    \qquad
    F_0=F_0\,,
    \qquad 
    F_2^{\mathrm{bg}}=e^a\Upsilon_a^-\,,
    \qquad
    F_4^{\mathrm{bg}}=f_\rho\Psi_+^\rho\,,
    \label{eq:iia-background-fluxes}
\end{equation}
in the flux units used throughout the paper. The internal gauge-invariant field strengths then take the form
\begin{align}
    H_3^{\mathrm{int}}
    &=
    \left(h^\rho+b^aN_a{}^\rho\right)\Phi_\rho^-\,,
    \label{eq:iia-H3-internal}\\
    \widetilde F_2^{\mathrm{int}}
    &=
    \left(e^a+F_0b^a\right)\Upsilon_a^-\,,
    \label{eq:iia-F2-internal}\\
    \widetilde F_4^{\mathrm{int}}
    &=
    \left[
        f_\rho+c^iM_{i\rho}
        +\kappa_{ab\rho}b^a
        \left(e^b+\frac{1}{2}F_0b^b\right)
    \right]\Psi_+^\rho\,.
    \label{eq:iia-F4-internal}
\end{align}

The Bianchi identities on the full field strengths are
\begin{equation}
    \dd H_3=0\,,
    \qquad
    \dd\widetilde F_2=F_0H_3+j_{\mathrm{O6/D6}}\,,
    \qquad
    \dd\widetilde F_4=H_3\wedge\widetilde F_2\,,
    \qquad
    \dd\widetilde F_6
    =
    H_3\wedge\widetilde F_4+j_{\mathrm{O2/D2}}\,.
    \label{eq:iia-Bianchi-identities}
\end{equation}
Writing the net source currents as
\begin{equation}
    j_{\mathrm{O6/D6}}
    =
    q_6^\rho\Phi_\rho^-\,,
    \qquad
    j_{\mathrm{O2/D2}}
    =
    q_2\,\widetilde{\mathrm{dvol}}_7\,,
    \label{eq:iia-source-expansions}
\end{equation}
where $\int_X\widetilde{\mathrm{dvol}}_7=1$, the axion-dependent terms in the O2 tadpole cancel by the preceding Bianchi identities, so its retained component can be written entirely in terms of the background fluxes. Upon integration, we find the following tadpole conditions:
\begin{align}
    h^\rho M_{i \rho}
    &=
    0\,,
    \label{eq:iia-H-bianchi}\\
    e^aN_a{}^\rho-F_0h^\rho
    &=
    q_6^\rho\,,
    \label{eq:iia-F2-bianchi}\\
    f_\rho N_a {}^{\rho}
    &=
   - \kappa_{ab\rho}e^b h^\rho\,,
    \label{eq:iia-F4-bianchi}\\
    -h^\rho f_\rho
    &=
    q_2\,.
    \label{eq:iia-F6-bianchi}
\end{align}
Note that the identity $\kappa_{ab\rho}q_6^\rho=0$ has been used for \eqref{eq:iia-F6-bianchi}, which follows from taking the Poincare dual of $\int j_{O6/D6}\wedge B_2$ and noting that $B_2$ is $\sigma$-odd and the cycle wrapped by $j_{O6/D6}$ is $\sigma$-fixed. 

The complete off-shell real superpotential can be written in components as
\begin{align}
    P_{\mathrm{IIA}}
    &=
    \underbrace{
    \frac{e^{-\phi/2}}{4\cV^2}
    \psi_\rho
    \left(h^\rho+b^aN_a{}^\rho\right)
    }_{P_H}
    \notag\\
    &\quad
    +
    \underbrace{
    \frac{e^{\phi/4}}{4\cV^2}
    s^\rho
    \left[
        f_\rho+c^iM_{i\rho}
        +\kappa_{ab\rho}b^a
        \left(e^b+\frac{1}{2}F_0b^b\right)
    \right]
    }_{P_{F_4}}
    +
    \underbrace{
    \frac{e^{5\phi/4}}{4\cV}F_0
    }_{P_{F_0}}\,.
    \label{eq:iia-component-superpotential}
\end{align}
The notation $P_H$, $P_{F_4}$, and $P_{F_0}$ refers to the three physical sectors explicitly displayed in \eqref{eq:iia-component-superpotential}. The second sector includes the axion-dressed $\widetilde F_4^{\mathrm{int}}$ flux and hence also the information about $\widetilde F_2^{\mathrm{int}}$. The source terms in the scalar potential are reproduced by cross terms among these three sectors after imposing the tadpole conditions~\cite{Farakos:2020phe,Aikot:2026tzn}.

The supersymmetry conditions require the following derivatives of $P_{\mathrm{IIA}}$, taken with respect to $\{\phi,s^\rho,b^a,c^i\}$, to vanish:
\begin{align}
    \begin{split}
      \partial_\phi P_{\mathrm{IIA}}&=-\frac{e^{-\phi/2}}{8\mathcal V}G_{\rho\sigma}^-(h^\rho+b^aN_a{}^\rho)s^\sigma\\
      &\quad+\frac{e^{\phi/4}}{16\mathcal V^2}s^\rho\left(f_\rho+c^iM_{i\rho}+\kappa_{ab\rho}b^a\left(e^b+\frac 12F_0b^b\right)\right)\\
      &\quad+\frac{5}{16\mathcal V}e^{5\phi/4}F_0\,,
    \end{split}\label{IIAEomDilaton}\\
    \begin{split}
        \partial_\rho P_{\mathrm{IIA}}&=-\frac{e^{-\phi/2}}{4\mathcal V}\left(\frac 13G^-_{\rho\sigma}G^-_{\tau\pi}s^\sigma s^\tau(h^\pi+b^aN_a{}^\pi)+G^-_{\rho\sigma}(h^\sigma+b^aN_a{}^\sigma)\right)\\
        &\quad+\frac{e^{\phi/4}}{4\mathcal V^2}\left(f_\rho+c^iM_{i\rho}+\kappa_{ab\rho}b^a\left(e^b+\frac 12F_0b^b\right)\right.\\
        &\qquad\qquad\left.-\frac 23G^-_{\rho\sigma}s^\sigma s^\tau\left(f_\tau+c^iM_{i\tau}+\kappa_{ab\tau}b^a\left(e^b+\frac 12F_0b^b\right)\right)\right)\\
        &\quad -\frac{e^{5\phi/4}}{12\mathcal V}G^-_{\rho\sigma}s^\sigma F_0\,,
    \end{split}\label{IIAEomMetric}\\
    \partial_aP_{\mathrm{IIA}}&=\frac{e^{-\phi/2}}{4\mathcal V}N_a{}^\rho G_{\rho\sigma}^-s^\sigma+\frac{e^{\phi/4}}{4\mathcal V^2}\kappa_{ab\rho}(e^b+F_0b^b)s^\rho\,,\label{IIAEomNS}\\
    \partial_iP_{\mathrm{IIA}}&=\frac{e^{\phi/4}}{4\mathcal V^2}M_{i\rho}s^\rho\label{IIAEomR}\,.
\end{align}
It is useful to give the last two supersymmetry equations a geometric interpretation. The orientifold projection sets $\tau_0=0$ and $\tau_1=0$ because our construction does not retain one-forms. On-shell, \eqref{IIAEomR} is equivalent to $\dd\Phi=0$, so the retained $\tau_3$ component vanishes at a supersymmetric vacuum and leaves only $\tau_2$ free. 

Writing $\tau_2=\tau_2^a\Upsilon_a^-$, the relation $\dd\Psi=\tau_2\wedge\Phi$ gives
\begin{equation}
    N_a{}^\rho G^-_{\rho\sigma}s^\sigma=G^-_{ab}\tau_2^b\,,
    \label{eq:iia-tau2-components}
\end{equation}
relating $\partial_aP_H$ to $\tau_2$. Thus, \eqref{IIAEomNS} can be written as 
\begin{equation}
    \tau_2=e^{3\phi/4}\widetilde F_2^{\mathrm{int}}\,.
    \label{eq:iia-tau2-F2-susy}
\end{equation}
In other words, supersymmetry does not require $\tau_2$ or $\widetilde F_2^{\mathrm{int}}$ to vanish separately; rather, it relates the intrinsic torsion to the axion-dressed two-form flux~\cite{VanHemelryck:2022ynr}.

\subsection{Universal flux balance and conformal dimension bounds}
\label{subsec:iia-flux-balance}

Two supersymmetry equations can be parameterized in terms of $\{P_{H,\star},P_{F_4,\star},P_{F_0,\star}\}$ without knowing the explicit metric and, consequently, $\psi_\rho(s)$. First, the dilaton equation gives
\begin{equation}
    0
    =
    -\frac{1}{2}P_{H,\star}
    +\frac{1}{4}P_{F_4,\star}
    +\frac{5}{4}P_{F_0,\star}\,.
    \label{eq:iia-dilaton-balance}
\end{equation}
Second, the contraction of the geometric supersymmetry equations with $s^\rho$ gives
\begin{equation}
    0
    =
    -\frac{10}{3}P_{H,\star}
    -\frac{11}{3}P_{F_4,\star}
    -\frac{7}{3}P_{F_0,\star}\,.
    \label{eq:iia-volume-balance}
\end{equation}
Solving \eqref{eq:iia-dilaton-balance} and \eqref{eq:iia-volume-balance} gives the universal on-shell relation
\begin{equation}
    P_{H,\star}:P_{F_4,\star}:P_{F_0,\star}
    =
    3:-4:2\,.
    \label{eq:iia-universal-flux-balance}
\end{equation}
Thus, the three sectors cannot be chosen independently at a supersymmetric vacuum. The on-shell superpotential and vacuum energy are consequently
\begin{equation}
    P_\star=\frac 12P_{F_0,\star}\,,\qquad V_\star=-P^2_{F_0,\star}\,.\label{eq:iia-on-shell-superpotential-energy}
\end{equation}
Every finite supersymmetric vacuum on the massive branch $F_0\neq0$ is therefore AdS$_3$, with $L^{-2}_\text{AdS}=P^2_{F_0,\star}$. If $F_0=0$, the same two equations require $P_{H,\star}=P_{F_4,\star}=0$, and the supersymmetric vacuum is Minkowski; we study this scenario in more detail in subsection \ref{subsec-iia-minkowski} and take $F_0\ne 0$ for the present discussion.

Given the universal weights of the superpotential terms, we can put bounds on the eigenvalues of the Hessian and thus the conformal dimensions, via the Cauchy interlacing theorem. We restrict ourselves to the two-dimensional subspace of moduli space that is spanned by the dilaton and the internal volume. Consider the vector fields 
\begin{equation}
    D^M_1=\delta^M{}_\phi\,,\qquad D^M_2= \delta^M{}_\rho s^\rho,
\end{equation}
so that $D^M_1\partial_M=\partial_\phi$ and $D^M_2\partial_M=s^\rho\partial_\rho$. We denote $\partial_\phi P_A=\omega_1^{(A)}P_A$ and $s^\rho\partial_\rho P_A=\omega_2^{(A)}P_A$, where $A$ indexes the three terms of the superpotential. The field space metric on this restricted subspace is obtained from \eqref{scalar-field-metric}, which is projected to
\begin{equation}
    \mathcal G_{pq}\equiv D^M_pD^N_q\mathcal G_{MN}=\begin{pmatrix}\frac 14&0\\0&\frac{28}9\end{pmatrix}\,.
\end{equation}
The superpotential Hessian $\mathcal H$ cleanly projects to this subspace:
\begin{equation}
    \mathcal H^p{}_{q}\equiv\left.\mathcal G^{pr}D^M_rD^N_q\nabla_M\partial_NP\right|_\star=\mathcal G^{pr}\sum_A\omega_r^{(A)}\omega_q^{(A)}P_{A,\star}=P_{\star}\begin{pmatrix}\frac{29}2&\frac{34}{3}\\\frac{51}{56}&-\frac{43}{14}\end{pmatrix}.
\end{equation}
The eigenvalues $\mu_\pm$ of the above matrix are 
\begin{equation}
    \frac{\mu_\pm}{2P_\star} = \frac{20 \pm 4 \sqrt{67}}{7}\,,\qquad \frac{\mu_-}{2P_\star} \approx -1.8\,,\quad  \frac{\mu_+}{2P_\star} \approx 7.5\,.
\end{equation}
Given that these values correspond to a subspace of the moduli space, they constrain the actual minimal and maximal eigenvalues of the full self-adjoint operator in \eqref{eq:general-superpotential-Hessian}. Choosing the irrelevant overall sign such that $P_\star>0$ we have
\begin{equation}
    \lambda_\text{min}\le\mu_-<\mu_+\le\lambda_\text{max}\,.
\end{equation}
The above inequality allows us to put a lower bound on the largest mass/conformal dimension in AdS units by \eqref{eq:general-normalized-Hessian-mass}:
\begin{equation}
    m_{\text{max}}^2L_{\mathrm{AdS}}^2 \geq \left(\frac{\mu_+}{2P_\star}-1\right)^2-1 \approx 41.7\,;
\end{equation}
the lower restricted eigenvalue $\mu_-$ does not directly constrain the smallest conformal dimension $\Delta_\text{min}$ but rather the conformal dimension corresponding to the smallest eigenvalue of the superpotential Hessian.

The corresponding bound on the scaling dimension of the dual conformal operator is obtained via \eqref{eq:general-conformal-dimension} and is given by
\begin{equation}
    \Delta_\text{max}\ge\Delta(\mu_+)= 1+\left|\frac{\mu_+}{2P_\star}-1\right| \approx 7.53\,.
\end{equation}
This is in agreement with the integer conformal dimensions of \cite{Miao:2025rgf, Farakos:2025bwf}. 

\subsection{Moduli stabilization and rank constraints}
\label{subsec:iia-axions}

\paragraph{$B_2$ axions.} The $B_2$ axions are fixed, for given values of $\phi$ and $s^\rho$, by \eqref{eq:iia-tau2-F2-susy}:
\begin{equation}
    b^a=\frac 1{F_0}\left(e^{-3\phi/4}(G^{-1}_-)^{ab}N_b{}^\rho G^-_{\rho\sigma}s^\sigma-e^a\right)\,.
    \label{IIAbSolution}
\end{equation}
For $N_a{}^\rho=0$ this reduces to the flux-only value $b^a=-e^a/F_0$.

\paragraph{$C_3$ axions.} From \eqref{eq:iia-potential-expansions}, we see that the $C_3$ axions are charged under the vectors $A^\alpha$, whenever the even non-closure matrix $N_\alpha{}^i$ is nonzero. Their covariant derivatives and gauge transformations are
\begin{equation}
    Dc^i=\dd c^i-N_\alpha{}^iA^\alpha\,, \qquad \delta c^i=N_\alpha{}^i\lambda^\alpha\,, \qquad \delta A^\alpha=\dd\lambda^\alpha\,.
    \label{eq:iia-c-axion-gauging}
\end{equation}
The superpotential is gauge invariant under \eqref{eq:iia-c-axion-gauging}. Consequently, $\operatorname{rank}N_\alpha{}^i$ combinations of $c^i$ are eaten. The only dependence of the superpotential on the remaining $c^i$ occurs through $c^iM_{i\rho}$, giving a necessary condition for all physical $C_3$ axions to be stabilized perturbatively:
\begin{align}
    \operatorname{rank}M_{i\rho}=b_3^+-\operatorname{rank}N_\alpha{}^i\,.
    \label{IIA caxion stabilization}
\end{align}
When this condition holds, the physical $C_3$ axions can be fixed by \eqref{IIAEomMetric}, which can be written in the form
\begin{equation}
    \left(\delta_\rho{}^\tau-\tfrac23G^-_{\rho\sigma}s^\sigma s^\tau\right)\widetilde F^{\mathrm{int}}_{4,\tau}
    =-4\mathcal V^2e^{-\phi/4}\left(\partial_\rho P_H+\partial_\rho P_{F_0}\right)\,.
\end{equation}
The operator acting on $\widetilde F^\text{int}_4$ has an inverse given by the Sherman--Morrison formula, $\delta_\rho{}^\tau-\tfrac2{11}G^-_{\rho\sigma}s^\sigma s^\tau$, allowing us to write
\begin{equation}
    \widetilde F^{\mathrm{int}}_{4,\rho}
    =\cV e^{-3\phi/4}G^-_{\rho\sigma}\!\left(h^\sigma+b^aN_a{}^\sigma\right)
    -\tfrac1{11}\!\left[3e^{-3\phi/4}\psi_\tau\!\left(h^\tau+b^aN_a{}^\tau\right)+\cV e^{\phi}F_0\right]G^-_{\rho\sigma}s^\sigma\,.
    \label{eq:iia-red-V}
\end{equation}
Upon contracting with $(G_-^{-1})^{\rho\lambda}M_{j\lambda}$, both terms in the right-hand side disappear by \eqref{eq:iia-H-bianchi}, \eqref{nilpotency}, and \eqref{IIAEomR}. What remains is the statement that the dressed $F_4$ flux is orthogonal, in the metric $G^-$, to the exact forms $\dd\Phi^+_j$,
\begin{equation}
    M_{j\lambda}(G_-^{-1})^{\lambda\rho}\widetilde F^{\mathrm{int}}_{4,\rho}=0\,.
\end{equation}
This is linear in $c^i$ and is solved by
\begin{equation}
    c^i=-\left[\left(M_{k\rho}(G_-^{-1})^{\rho\sigma}M_{l\sigma}\right)^{-1}\right]^{ij}M_{j\tau}(G_-^{-1})^{\tau\lambda}\left(f_\lambda+\kappa_{ab\lambda}b^a\left(e^b+\tfrac12F_0b^b\right)\right),
    \label{eq:iia-c-solution}
\end{equation}
with the inverse taken on the image of $M_{i\rho}$.

\paragraph{Geometric moduli and dilaton.} The geometric moduli are to be stabilized using \eqref{IIAEomR} and the remaining equations in \eqref{IIAEomMetric}, the former of which readily tells us that $s^\rho\in\ker(M_{i\rho})$. Since the vector $s^\rho$ must be nonzero, a necessary condition for a solution is $\dim\ker(M_{i\rho})\geq1$, or equivalently,
\begin{align}
    \operatorname{rank}M_{i\rho}<b_3^-\,.
    \label{IIAm<b3}
\end{align}
For geometries satisfying \eqref{IIAm<b3}, let $m$ be the dimension of $\ker(M_{i\rho})$ and let $v_r^\rho$, where $r=1,\ldots,m$, be a fixed basis of this kernel. We parametrize the geometric moduli as $s^\rho=t^r v_r^\rho$, where the $m$ real coefficients $t^r$ are to be determined. After eliminating $b^a$ and contracting with the vectors $v^\rho_r$, \eqref{IIAEomMetric} is written as
\begin{align}
    0={}&-v_r^\rho G^-_{\rho\sigma}\left(h^\sigma-\frac{N_a{}^\sigma e^a}{F_0}\right)\nonumber\\
    &-\frac{e^{-3\phi/4}}{F_0}v_r^\rho G^-_{\rho\sigma}N_a{}^\sigma(G_-^{-1})^{ab}N_b{}^\lambda G^-_{\lambda\tau}s^\tau\nonumber\\
    &+\frac{e^{3\phi/4}}{\mathcal V}v_r^\rho\left(f_\rho-\frac{1}{2F_0}\kappa_{ab\rho}e^ae^b\right)\nonumber\\
    &+\frac{e^{-3\phi/4}}{2F_0\mathcal V}v_r^\rho\kappa_{ab\rho}(G_-^{-1})^{ac}(G_-^{-1})^{bd}\left(N_c{}^\lambda G^-_{\lambda\mu}s^\mu\right)\left(N_d{}^\kappa G^-_{\kappa\nu}s^\nu\right)\nonumber\\
    &+\frac12e^{7\phi/4}F_0v_r^\rho G^-_{\rho\sigma}s^\sigma\,.
    \label{IIAKernelSaxion}
\end{align}
The $c^i$ dependence, which enters only through $c^iM_{i\rho}$, has dropped out. These $m$ equations \eqref{IIAKernelSaxion} with \eqref{IIAEomDilaton} generically fix $\phi$ and the $m$ coefficients $t^r$, and hence the geometric moduli~$s^\rho$. 

\subsection{\texorpdfstring{Universal $B_2$-axion integer dimension}{Universal B2-axion integer dimension}}
\label{subsec:iia-universal-b2-spectrum}

The $B_2$-axion masses can be determined exactly when the retained orientifold-odd two-forms are closed, $N_a{}^\rho=0$, which implies $\tau_2=0$ by \eqref{eq:iia-tau2-components}. Then, the $B_2$ axions do not mix at quadratic order with the dilaton, the geometric moduli, or the RR axions, whereas their own Hessian block is
\begin{equation}
    \left.\nabla_{b^a}\partial_{b^b}P_{\mathrm{IIA}}\right|_\star
    =
    \frac{e^{\phi_\star/4}F_0}{4\cV_\star^2}
    s_\star^\rho\kappa_{ab\rho}
    =
    -\frac{e^{\phi_\star/4}F_0}{4\cV_\star}
    G_{ab,\star}^-\,.
    \label{eq:iia-b-superpotential-Hessian}
\end{equation}
Raising one index with the physical metric yields a multiple of the identity, 
\begin{align}
    \mathcal{H}^a{}_b
    &=
    \left.
     4 e^\phi (G_-^{-1})^{ac}
    \nabla_{b^c}\partial_{b^b}P_{\mathrm{IIA}}
    \right|_\star
    \notag\\
    &=
    -\frac{e^{5\phi_\star/4}F_0}{\cV_\star}
    \delta^a{}_b
    =
    -4P_{F_0,\star}\delta^a{}_b
    =
    -8P_\star\delta^a{}_b\,.
    \label{eq:iia-b-normalized-superpotential-Hessian}
\end{align}
Every one of the $b_2^-$ eigenvalues therefore obeys
\begin{equation}
    \frac{\lambda_b}{2P_\star}
    =
    -4\,,\qquad m_b^2L_{\mathrm{AdS}}^2=24\,,
    \qquad
    \Delta_b=6.
    \label{eq:iia-universal-b-mass-dimension}
\end{equation}
In other words, in this torsionless scenario, the entire $B_2$-axion block is degenerate and independent of the detailed $G_2$ metric, the intersection numbers, and the flux values. We emphasize that, unlike the integer dimensions that arise only asymptotically in several flux-scaling examples, \eqref{eq:iia-universal-b-mass-dimension} holds at every vacuum. 

The exact dimension $\Delta_b=6$ in \eqref{eq:iia-universal-b-mass-dimension} sits on the list of AdS$_3$ values admitting a polynomial shift symmetry, 
\begin{equation}
    m^2L_{\mathrm{AdS}}^2=k(k+2),\qquad \Delta=2+k,
\end{equation}
with $k$ a non-negative integer~\cite{Hinterbichler:2014cwa,Bonifacio:2018zex}, at $k=4$; the benchmark spectrum \cite{Miao:2025rgf, Farakos:2025bwf} likewise lies on this list, at $k=\{6,2,2,2,2,0,0,0\}$. This pattern has been proposed as the origin of the integer dimensions of flux vacua~\cite{Apers:2022vfp} and has been checked against large families of three-dimensional orientifold vacua~\cite{Arboleya:2025ocb}. The difference is that the benchmark values are reached only in the large-flux limit, whereas \eqref{eq:iia-universal-b-mass-dimension} holds at finite flux, so the free scalar theory sits exactly at a polynomial-shift-symmetric mass.

The integer conformal dimension admits a super-extremal arrangement of the conformal dimensions, namely $(6,2,2)$. In an AdS effective theory with finitely many fields and a central charge $c$, a consistency argument in the dual CFT requires the cubic coupling of scalars whose dimensions satisfy 
\begin{equation}
    \Delta_i=\Delta_j+\Delta_k+2n\,,\qquad n\in\mathbb N
\end{equation}
to vanish, since the dual single-particle operator cannot be defined with large-$c$ factorization~\cite{Bobev:2025yxp}. Dimensions that are asymptotically integer evade this requirement when their deviation vanishes more slowly than $c^{-1/2}$, since it then regulates the divergence of the three-point Witten diagram; for the family of~\cite{Farakos:2025bwf} it scales as $c^{-1/8}$. Exact dimensions such as \eqref{eq:iia-universal-b-mass-dimension} have no such protection. Its coupling nevertheless vanishes by symmetry. With $N_a{}^\rho=0$, the $B_2$ axions are perturbatively stabilized at $b_\star^a=-e^a/F_0$. Writing $b^a=b_\star^a+\delta^a$, the only $b$-dependence of \eqref{eq:iia-component-superpotential} becomes $\kappa_{ab\rho}b^a(e^b+\tfrac12F_0b^b)=\tfrac12F_0\kappa_{ab\rho}(\delta^a\delta^b-b_\star^ab_\star^b)$, with the Lagrangian invariant under $\delta^a\to-\delta^a$. Every cubic coupling with a single $B_2$ fluctuation therefore vanishes and the holographic requirement is satisfied.

\subsection{Large--flux limits and isotropic scaling}
\label{subsec:iia-large-f4}

We next consider the large-flux scenarios in which one background flux vector becomes large while the source content is kept fixed. We first consider $f_\rho=nf^{(0)}_\rho+\mathcal O(1)$, where $n\to\infty$ and $f^{(0)}_\rho\in\mathbb Z^{b^+_4}$, while all other fluxes and the source content are kept fixed. The relevant Bianchi identities \eqref{eq:iia-F4-bianchi} and \eqref{eq:iia-F6-bianchi} take the form
\begin{equation}
    n f_\rho^{(0)}N_a{}^\rho+\mathcal O(1)=-\kappa_{ab\rho}e^bh^\rho\,,\qquad -n h^\rho f_\rho^{(0)}+\mathcal O(1)=q_2\,.
\end{equation}
The right-hand sides remain of order one, so the coefficients of the order-$n$ terms on the left-hand sides must vanish. Therefore
\begin{equation}
    f_\rho^{(0)}N_a{}^\rho=0\,,\qquad h^\rho f_\rho^{(0)}=0\,.
    \label{eq:iia-large-F4-kernel-conditions}
\end{equation}
Thus, the unbounded $F_4$ flux must be both closed in the truncated form complex and orthogonal to the background $H_3$ flux. However, the kernel count is not yet the number of physically independent scaling parameters. Equation \eqref{eq:iia-F4-internal} is invariant under the simultaneous change of flux representative and RR axion,
\begin{equation}
    f_\rho
    \longrightarrow
    f_\rho+k^iM_{i\rho}\,,
    \qquad
    c^i
    \longrightarrow
    c^i-k^i\,.
    \label{eq:iia-F4-axion-representative-shift}
\end{equation}
The complete superpotential and all gauge-invariant field strengths are unchanged. Moreover, any $f_\rho\in\operatorname{Im}\bigl((M_{i\rho})^T\bigr)$ automatically satisfies \eqref{eq:iia-large-F4-kernel-conditions} by \eqref{nilpotency} and
\eqref{eq:iia-H-bianchi}. Generically, these conditions are independent, and the number of physically inequivalent scaling directions is $b_3^- - \operatorname{rank}N_a{}^\rho -1-\operatorname{rank}M_{i\rho}$. Such large-$F_4$ scenarios have demonstrated parametric control for studying scale separation in~\cite{Farakos:2020phe}.

A putative large $F_2$ limit does not give an independent scaling direction in massive type IIA. To see this, write $e^a=\ell e^a_{(0)}+\mathcal O(1)$ with $\ell\to\infty$. The Bianchi identities \eqref{eq:iia-F2-bianchi} and \eqref{eq:iia-F4-bianchi} require at leading order
\begin{equation}
    e^a_{(0)}N_a{}^\rho=0\,,\qquad \kappa_{ab\rho}e^b_{(0)}h^\rho=0\,.
    \label{eq:iia-large-F2-kappa-kernel}
\end{equation}
The first condition implies that the growing flux lies along a closed integral two-form and can therefore participate in a large $B_2$ gauge transformation. Furthermore, \eqref{stokes} together with \eqref{eq:iia-large-F2-kappa-kernel} implies that the induced vector $\kappa_{ab\rho}e^a_{(0)}e^b_{(0)}$ obeys precisely the two large-$F_4$ conditions in \eqref{eq:iia-large-F4-kernel-conditions}. For $\lambda^a\in\ker N_a{}^\rho\cap\mathbb Z^{b_2^-}$, invariance of the gauge-invariant field strengths requires
\begin{equation}
    b^a\longrightarrow b^a+\lambda^a\,,\qquad e^a\longrightarrow e^a-F_0\lambda^a\,,\qquad f_\rho\longrightarrow f_\rho-\kappa_{ab\rho}\lambda^ae^b+\frac12F_0\kappa_{ab\rho}\lambda^a\lambda^b\,.
    \label{eq:iia-F2-axion-representative-shift}
\end{equation}
Writing $\ell=F_0q+r$ with bounded $r$ and choosing $\lambda^a=q e^a_{(0)}$ removes the unbounded part of $e^a$. The combination
\begin{equation}
    f_\rho-\frac{1}{2F_0}\kappa_{ab\rho}e^ae^b
\end{equation}
is invariant under \eqref{eq:iia-F2-axion-representative-shift}. Hence, if this gauge-invariant flux combination grows along the putative large-$F_2$ family, the same family can be represented with bounded $e^a$ and parametrically large $F_4$ flux, generically of order $\ell^2$; if it does not grow, there is no physical large-flux parameter. Thus, there is no independent large-$F_2$ limit in massive type IIA. 

We now turn to the standard large-$F_4$ scaling of the torsion-free models~\cite{Farakos:2020phe}. Consider an isotropic solution with a fixed shape $u^\rho$ and assume that the axion-dressed fluxes entering $P_H$ and $P_{F_4}$ retain their standard leading scaling, so that $h^\rho+b^aN_a{}^\rho=\mathcal O(1)$ and the axion-dressed $F_4$ flux is of order $n$. Balancing the three superpotential sectors as required by \eqref{eq:iia-universal-flux-balance} then gives
\begin{equation}
    s^\rho\sim n^{21/16}u^\rho\,,\qquad e^\phi\sim n^{-3/4}\,,\qquad \cV_E=\cV\sim n^{49/16}\,,\qquad \cV_s=e^{7\phi/4}\cV_E\sim n^{7/4}\,.
    \label{eq:iia-large-F4-moduli-scaling}
\end{equation}
Each of $P_H$, $P_{F_4}$, and $P_{F_0}$ then scales as $n^{-4}$, and
\begin{equation}
    P_\star\sim n^{-4}\,,\qquad |V_\star|=4P_\star^2\sim n^{-8}\,.
    \label{eq:iia-large-F4-vacuum-scaling}
\end{equation}
For an isotropic internal metric, the lightest Kaluza--Klein mass in three-dimensional Einstein frame scales as
\begin{equation}
    m_{\mathrm{KK}}^2\sim\cV_E^{-16/7}\sim n^{-7}\,,\qquad \frac{L_{\mathrm{KK}}^2}{L_{\mathrm{AdS}}^2}=\frac{|V_\star|}{m_{\mathrm{KK}}^2}\sim n^{-1}\,.
    \label{eq:iia-large-F4-scale-separation}
\end{equation}
This reproduces the familiar weak-coupling, large-volume, scale-separated limit.

The supersymmetry relation \eqref{eq:iia-tau2-F2-susy} makes the assumption that curvature does not modify this leading scaling nontrivial on a $G_2$-structure space. Using \eqref{IIAbSolution}, the standard scaling \eqref{eq:iia-large-F4-moduli-scaling} gives
\begin{equation}
    e^a+F_0b^a=e^{-3\phi/4}(G_-^{-1})^{ab}N_b{}^\rho G^-_{\rho\sigma}s^\sigma\sim n
\end{equation}
for a fixed limiting shape satisfying $N_b{}^\rho G^-_{\rho\sigma}(u)u^\sigma\neq0$. For the supersymmetric configurations considered here, $R_7=-\frac12|\tau_2|^2$~\cite{2003math......5124B}. Hence the same generic scaling gives
\begin{equation}
    V_{\mathrm{curv}}^{(\tau_2)}=-\frac{1}{2\cV_E^3}\int_X\dd^7y\,\sqrt{g_7}\,R_7\sim n^{-7}\,,\qquad V_{\widetilde F_2}\sim n^{-7}\,.
\end{equation}
Both contributions are therefore of order the Kaluza--Klein mass squared in our gravitational normalization, rather than the desired vacuum energy $|V_\star|\sim n^{-8}$. Moreover, the resulting scaling of $b^a$ modifies the axion-dressed $H_3$ and $F_4$ fluxes, so the assumptions used to derive \eqref{eq:iia-large-F4-moduli-scaling} cease to be self-consistent. Thus, a generic fixed torsion deformation obstructs the standard large-$F_4$ hierarchy, in agreement with the general curvature obstruction for isotropic compactifications~\cite{Gautason:2015tig,Tringas:2025uyg}.

\subsection{Exact flat direction in supersymmetric Minkowski vacua}
\label{subsec-iia-minkowski}

A universal massless volume--dilaton direction was previously identified for type IIA compactifications on $G_2$-holonomy manifolds~\cite{Farakos:2020phe}. We now show that it persists for the more general $G_2$-structure compactifications considered here, including intrinsic torsion and the axionic sector.

At a finite supersymmetric Minkowski vacuum, $V_\star=0$, and equations \eqref{eq:iia-on-shell-superpotential-energy} and \eqref{eq:iia-universal-flux-balance} require $P_{F_0,\star}=P_{H,\star}=P_{F_4,\star}=0$. Combining the dilaton and volume weights displayed in \eqref{eq:iia-dilaton-balance} and \eqref{eq:iia-volume-balance}, we find that, for $F_0=0$, the two remaining superpotential terms scale identically under
\begin{equation}
s^\rho\longrightarrow e^{3t/7}s^\rho\,,\qquad \cV\longrightarrow e^t\cV\,,\qquad \phi\longrightarrow\phi+\frac{4}{21}t\,.
\end{equation}
Both $P_H$ and $P_{F_4}$ acquire the common factor $e^{-32t/21}$. Consequently, conditions $P_{\mathrm{IIA}}=0$ and $\partial_M P_{\mathrm{IIA}}=0$ remain satisfied along this trajectory, with all axions and shape ratios kept fixed. Thus every finite supersymmetric Minkowski solution belongs to a continuous one-parameter family, and one combination of the overall volume and dilaton is necessarily an exact classical flat direction.

\section{Type IIB and type I}
\label{sec:type-IIB-type-I}
In this section, we study supersymmetric vacua of the three-dimensional low-energy effective theory for type IIB flux compactifications on $G_2$-structure spaces with O5/O9-planes, together with their type I truncations, extending \cite{Aikot:2026tzn}. We derive the universal conformal dimensions from the universal supersymmetric dilaton--volume sector and the $B_2$-axion dimensions on a restricted singlet-torsion branch. We analyze the rank conditions for full moduli stabilization and show that every supersymmetric Minkowski vacuum retains an exact classical flat direction. 

\subsection{Tadpole conditions and equations of motion}
\label{subsec:iib-setup}

The O5/O9 orientifold projection and its action on the $G_2$ forms are given by~\cite{Angelantonj:2002ct,GrimmLouisIIB0403067,Aikot:2026tzn}
\begin{equation}
    \mathcal{O}
    \equiv
    \Omega_p\sigma\,,
    \qquad
    \sigma^*\Phi=+\Phi\,,
    \qquad
    \sigma^*\Psi=+\Psi\,,
    \label{eq:iib-orientifold-action}
\end{equation}
where $\sigma$ is the involutive isometry of the target space and $\Omega_p$ is the worldsheet parity. 

Upon projection, we retrieve the same parity-restricted subspaces as in \eqref{Projected-basis}, namely 
\begin{equation}
    \begin{aligned}
    \Upsilon_A &\to \{\Upsilon_\alpha^+, \Upsilon_a^-\}\,,  &\Phi_I &\to \{\Phi^+_i, \Phi^\rho_-\}\,,\\
    \Psi^I&\to\{\Psi^i_+,\Psi^-_\rho\}\,, &T^A&\to\{T^\alpha_+,T^a_-\}\,,
    \end{aligned}
\end{equation}
where the indices have the same range as section~\ref{subsec:iia-setup}. The even and odd subspaces of the torsion matrices, the triple-intersection numbers, and the 2-, 3-, and 4-form metrics are denoted by Greek and Roman indices, similar to Section~\ref{subsec:iia-setup}. The equations \eqref{orthonormality}--\eqref{eq:metric-intersection-identity} hold in their respective sectors. 

The $G_2$ structure and its dual are then expanded in even three- and four-forms.
\begin{equation}
    \Phi=s^i\Phi^+_i,\qquad\Psi=\psi_i(s)\Psi^i_+,\qquad\psi_i(s)=3\partial_i\mathcal V.\label{g2-structure-IIB}
\end{equation}
The surviving ten-dimensional potentials are
\begin{equation}
    B_2=b^a\Upsilon_a^-\,,\qquad C_2=c_2^\alpha\Upsilon^+_\alpha\,,\qquad C_4=c^\rho_4\Psi^-_\rho+A_\rho\wedge\Phi_-^\rho+\ldots\,.\label{IIB-Flux-Expansion}
\end{equation}
The omitted terms are non-dynamical in three dimensions. Although the self-duality of $\widetilde F_5$ relates the scalar and vector descriptions in \eqref{IIB-Flux-Expansion}, it is useful to retain both temporarily in order to identify the gauge-inequivalent scalar directions.

We expand the quantized background Page fluxes as
\begin{equation}
    H^\text{bg}_3=h_\rho\Phi^\rho_-\,,\qquad F^\text{bg}_3=f^i\Phi^+_i\,,\qquad F^\text{bg}_5=e_aT^a_-\,,\qquad F^\text{bg}_7=f_7\widetilde{\text{dvol}_7}\,.
\end{equation}
Using the Page-flux convention of~\cite{Aikot:2026tzn}, the internal gauge-invariant field strengths are
\begin{align}
H_3^{\mathrm{int}}
&=
\left(
h_\rho+b^aN_{a\rho}
\right)\Phi^{\rho}_-,
\\[2mm]
\widetilde F_3^{\mathrm{int}}
&=
\left(
f^i+c_2^\alpha N_\alpha{}^i
\right)\Phi_i^+,
\\[2mm]
\widetilde F_5^{\mathrm{int}}
&=
\left[
e_a
+c_4^\rho Q_{\rho a}
+\kappa_{abi}b^bf^i
-\kappa_{a\alpha}{}^\rho c_2^\alpha
\left(
h_\rho+b^bN_{b\rho}
\right)
\right]T^{a}_-,
\label{iib: F_5}
\\[2mm]
\widetilde F_7^{\mathrm{int}}
&=
\left[
f_7
+b^ae_a
+\frac12\kappa_{abi}b^ab^bf^i
-c_4^\rho
\left(
h_\rho+b^aN_{a\rho}
\right)
\right]
\widetilde{\text{dvol}}_7.
\label{iib: F_7}
\end{align}

The Bianchi identities are 
\begin{equation}
    \dd H_3=0\,, \qquad 
    \dd\widetilde F_3=j_{\mathrm{O5/D5}}\,, \qquad 
    \dd\widetilde F_5=H_3\wedge\widetilde F_3\,, \qquad 
    \dd\widetilde F_7 = H_3\wedge\widetilde F_5\,.
\end{equation} 
Writing $j_{\mathrm{O5/D5}} = q_{5,j}\Psi_+^j$, the cohomological constraints in the retained three- and four-form sector are
\begin{align}
    h_\rho M^{\rho\sigma}
    &=
    0\,,
    \label{eq:iib-H3-Bianchi-components}\\
    f^iM_{ij}
    &=
    q_{5,j}\,,
    \label{eq:iib-O5-tadpole-components}
\end{align}
with the omitted two Bianchi identities trivially satisfied. In the unwarped two-derivative reduction with mutually calibrated O5/D5 sources, the torsion-free tadpole condition $q_{5,j}=0$ cancels their net tension contribution. The remaining potential $V=V_{H_3}+V_{F_3}+V_{F_5}+V_{F_7}$ is a sum of nonnegative flux norms. Under $s^i\to\lambda s^i$, at fixed dilaton, shape, axions, and flux representatives,
\begin{equation}
    \lambda\partial_\lambda V
    =-\frac{20}{3}\left(V_{H_3}+V_{F_3}\right)-8V_{F_5}-\frac{28}{3}V_{F_7}\leq0\,.
\end{equation}
At a finite critical point the derivative must vanish, forcing all retained gauge-invariant fluxes to vanish. Hence neither AdS nor dS nor nontrivial Minkowski critical points exist in this reduction; only a fluxless, source-cancelled Minkowski locus remains, with flat volume and dilaton.

The complete off-shell real superpotential is
\begin{align}
    P_{\mathrm{IIB}}
    &=
    \underbrace{
    \frac{e^{-\phi/2}}{4\cV^2}
    \left[
        f_7
        +
        b^a e_a
        +
        \frac{1}{2}\kappa_{abi}b^a b^b f^i
        -
        c_4^\rho
        \left(h_\rho+b^aN_{a\rho}\right)
    \right]
    }_{P_{F_7}}
    \notag\\
    &\quad
    \underbrace{
    -
    \frac{e^{\phi/2}}{4\cV^2}
    \psi_i(s)
    \left(f^i+c_2^\alpha N_\alpha{}^i\right)
    }_{P_{F_3}}
    +
    \underbrace{
    \frac{1}{8\cV^2}
    s^iM_{ij}s^j
    }_{P_{\mathrm{geom}}}\,.
    \label{eq:iib-component-superpotential}
\end{align}
Although there is no separate $P_{F_5}$ term, the five-form flux enters $P_{F_7}$ through $b^ae_a$, and its contribution to the scalar potential is generated by the axion derivatives of the superpotential. The self-dual five-form is treated in the standard pseudo-action formalism and its duality condition is imposed after variation. 

The supersymmetry conditions require the following derivatives of $P_{\mathrm{IIB}}$, taken with respect to $\{\phi,s^i,b^a,c_2^\alpha,c_4^\rho\}$, respectively, to vanish:
\begin{align}
    \begin{split}
        \partial_\phi P_{\mathrm{IIB}}
        &=
        -\frac{e^{-\phi/2}}{8\mathcal V^2}
        \left(
            f_7
            +
            b^a\left(
                e_a+\frac{1}{2}\kappa_{abi}b^bf^i
            \right)
            -
        c_4^\rho
        \left(h_\rho+b^aN_{a\rho}\right)
        \right)
        \\
        &\quad
        -\frac{e^{\phi/2}}{8\mathcal V}
        \left(
            G_{ij}^+f^is^j
            +
            c_2^\alpha G^+_{ij}N_\alpha{}^i s^j
        \right)\,,
    \end{split}
    \label{IIBEomDilaton}
    \\
    \begin{split}
        \partial_iP_{\mathrm{IIB}}
        &=
        -\frac{e^{-\phi/2}}{6\mathcal V^2}
        G^+_{ij}s^j
        \left(
            f_7
            +
            b^a\left(
                e_a+\frac{1}{2}\kappa_{abk}b^bf^k
            \right)
            -
            \left(
                h_\rho+b^aN_{a\rho}
            \right)c_4^\rho
        \right)
        \\
        &\quad
        +\frac{e^{\phi/2}}{4\mathcal V}
        \left(
            \frac{1}{3}G^+_{ij}G^+_{kl}s^js^k
            +
            G^+_{il}
        \right)
        \left(
            f^l+c_2^\alpha N_\alpha{}^l
        \right)
        +
        \frac{1}{4\mathcal V^2}
        \left(
            M_{ij}s^j
            -
            \frac{1}{3}G^+_{ij}M_{kl}s^js^ks^l
        \right)\,,
    \end{split}
    \label{IIBEomMetric}
    \\
    \partial_aP_{\mathrm{IIB}}
    &=
    \frac{e^{-\phi/2}}{4\mathcal V^2}
    \left(
        e_a
        +
        \kappa_{abi}b^bf^i
        -
        N_{a\rho}c_4^\rho
    \right)\,,
    \label{IIBEomNS}
    \\
    \partial_\alpha P_{\mathrm{IIB}}
    &=
    -\frac{e^{\phi/2}}{4\mathcal V}
    G^+_{ij}N_\alpha{}^i s^j\,,
    \label{IIBEomR2}
    \\
    \partial_\rho P_{\mathrm{IIB}}
    &=
    -\frac{e^{-\phi/2}}{4\mathcal V^2}
    \left(
        h_\rho+b^aN_{a\rho}
    \right)\,.
    \label{IIBEomR4}
\end{align}
As in type IIA, the axion supersymmetry equations have a geometric interpretation. While not projected out, the singlet torsion $\tau_0$, averaged over $X$, is precisely the geometric sector of \eqref{eq:iib-component-superpotential},
\begin{equation}
    \overline\tau_0=\frac{8}{7}\mathcal VP_\text{geom}.\label{eq:iib-tau0}
\end{equation}
The two-form torsion class is even, expanding as $\tau_2=\tau^\alpha_2\Upsilon^+_\alpha$. Given $\tau_1=0$, the even-parity version of \eqref{eq:iia-tau2-components} holds, and thus the $C_2$-axion equation \eqref{IIBEomR2} becomes 
\begin{equation}
    \partial_\alpha P_\text{IIB}=-\frac{e^{\phi/2}}{4\mathcal V}G^+_{\alpha\beta}\tau^\beta_2=0\implies\tau_2=0.
\end{equation}

\subsection{Universal supersymmetric flux balance}
\label{subsec:iib-flux-balance}

The two supersymmetry equations that balance the superpotential sectors $P_{F_7}$, $P_{F_3}$, $P_\text{geom}$ are
\begin{align}
    0&=-\frac{1}{2}P_{F_7,\star}+\frac{1}{2}P_{F_3,\star}\,,\label{eq:iib-dilaton-balance}\\
    0&=-\frac{14}{3}P_{F_7,\star}-\frac{10}{3}P_{F_3,\star}-\frac{8}{3}P_{\mathrm{geom},\star}\,.\label{eq:iib-volume-balance}
\end{align}
Solving \eqref{eq:iib-dilaton-balance} and \eqref{eq:iib-volume-balance}, we find
\begin{equation}
    P_{F_7,\star}:P_{F_3,\star}:P_{\mathrm{geom},\star}
    =
    1:1:-3\,.
    \label{eq:iib-universal-flux-balance}
\end{equation}
The on-shell superpotential and vacuum energy are consequently
\begin{equation}
    P_\star=\frac 13P_{\text{geom},\star}\,,\qquad V_\star=-\frac 49P^2_{\text{geom},\star}\,,\qquad L^{-2}_\text{AdS}=\frac 49P^2_{\text{geom},\star}.
\end{equation}
Two things are of note here. First, every finite supersymmetric vacuum with $s_\star^iM_{ij}s_\star^j\neq0$, or with a nonzero singlet torsion contribution, is AdS$_3$. Second, for an isotropic internal space every curvature radius scales as $\mathcal V^{1/7}$, so that $|V_\star|\sim\mathcal V^{-16/7}$. This is precisely the scaling of the three-dimensional Einstein-frame Kaluza--Klein mass squared used in \eqref{eq:iia-large-F4-scale-separation}, so the ratio $L_{\mathrm{KK}}^2/L_{\mathrm{AdS}}^2$ of \eqref{eq:general-scale-separation} is generically an order-one number that depends neither on the fluxes nor on the string coupling. We therefore find no isotropic large-$F_3$ limit with scale separation, in agreement with the general curvature obstruction of~\cite{Gautason:2015tig,Tringas:2025uyg}.

Nevertheless, an anisotropic internal geometry evades this obstruction. Then, the lightest Kaluza--Klein mass is set by the largest internal radius rather than by $\mathcal V^{1/7}$, as emphasized in subsection~\ref{subsec:general-control}, while the singlet torsion is controlled by ratios of radii and can be made parametrically small in string units without the total volume being large in an isotropic sense. Explicit scale-separated AdS$_3$ families of this type are known: supersymmetric type IIB vacua were constructed in~\cite{VanHemelryck:2025qok}, non-supersymmetric type IIB examples with parametrically controlled scale separation in~\cite{Arboleya:2024vnp}, type I vacua in~\cite{Miao:2025rgf}, and heterotic vacua in~\cite{Tringas:2025bwe}. In all such cases the lightest Kaluza--Klein eigenvalue has to be computed from the individual radii rather than inferred from the total volume~\cite{Coudarchet:2023mfs}. A class of supersymmetric scale-separated type IIB AdS$_3$ vacua also admits an interpretation as the near-horizon region of smeared D1--D5--KK5 configurations~\cite{Arboleya:2026irl}.

The mass and conformal dimension bounds of subsection~\ref{subsec:iia-flux-balance} apply without change. With $D_1=\partial_\phi$, $D_2=s^i\partial_i$, and the weights of the sectors $\{F_7,F_3,\mathrm{geom}\}$ as in \eqref{eq:iib-dilaton-balance} and \eqref{eq:iib-volume-balance}, the projected superpotential Hessian is
\begin{equation}
    \mathcal H^p{}_q
    =\mathcal G^{pr}\sum_A\omega^{(A)}_r\omega^{(A)}_q\,P_{A,\star}
    =P_\star\begin{pmatrix}-2&-\tfrac83\\[2pt]-\tfrac{3}{14}&-\tfrac{26}{7}\end{pmatrix}.
\end{equation}
Its eigenvalues and eigenvectors are
\begin{equation}
    \frac{\mu_-}{2P_\star}=-2\,,\quad
    D_-\equiv4\partial_\phi+3s^i\partial_i\,;
    \qquad
    \frac{\mu_+}{2P_\star}=-\frac67\,,\quad
    D_+\equiv-28\partial_\phi+3s^i\partial_i\,.
    \label{eq:iib-D-direction}
\end{equation}
Since both Ritz values are negative, the upper one gives no bound on $\Delta_{\max}$, while the lower one gives $\Delta(\lambda_{\min})\geq4$. Beyond bounding, we ask whether the Ritz vectors are exact eigenvectors, in which case $\Delta=4$ and $\Delta=20/7$ belong to the spectrum. In type IIA the volume--dilaton plane is not invariant in general~\cite{Farakos:2025bwf}, whereas in type IIB/type I the invariance conditions take a simple universal form, which we now derive.

For an eigenvector $D$ with eigenvalue $\mu$, let $\omega_A$ denote its weights on the sectors $\{F_7,F_3,\mathrm{geom}\}$. Homogeneity gives $D^N\partial_NP_{\mathrm{IIB}}=\sum_A\omega_AP_A$ identically, and at a supersymmetric point $\partial_MP_{\mathrm{IIB}}|_\star=0$ removes the Christoffel term and the derivative of $D^N$, so that
\begin{equation}    X_M\equiv\left.\left(\nabla_M\partial_NP_{\mathrm{IIB}}\right)D^N\right|_\star=\sum_A\omega_A\,\partial_MP_A\big|_\star\,.
\end{equation}
$D$ is an exact eigenvector of \eqref{eq:general-superpotential-Hessian} if and only if $X_M=\mu\,\mathcal G_{MN}D^N$. This holds along the plane spanned by the eigenvectors $D$ by construction. Since $\mathcal G_{MN}$ does not mix the dilaton, saxions and axions, and $\mathcal G_{ij}s^j=\tfrac49\psi_i/\cV$, the $\mathcal G$-orthogonal complement of the plane consists of the axions and the saxion directions with $\psi_i\delta s^i=0$. Hence $D$ is exact if and only if
\begin{equation}
    X_M\big|_{\text{axions}}=0\,,\qquad X_i\propto\psi_i\,.
    \label{eq:iib-ritz-exactness}
\end{equation}
The first condition holds for every $D$: $b^a$ and $c_4^\rho$ enter only $P_{F_7}$ and $c_2^\alpha$ only $P_{F_3}$, so their F-terms give $\partial_{\mathrm{ax}}P_A|_\star=0$ sector by sector. For the second, $\partial_iP_{F_7}\propto\psi_i$, whereas $\partial_iP_{\mathrm{geom}}\propto\psi_i$ if and only if $M_{ij}s^j\propto\psi_i$, that is, $\tau_3=0$. For $D_-$, it turns out that $X_M$ does not contain $\partial_MP_\text{geom}$, automatically satisfying both exactness conditions, while for $D_+$, the $\partial_MP_\text{geom}$ term is present and thus exactness requires $\tau_3=0$. We have, in summary,
\begin{equation}
    \begin{aligned}
        m_{D_-}^2L_{\mathrm{AdS}}^2&=8\,,&\qquad\Delta_{D_-}&=4\qquad&&\text{at every supersymmetric vacuum,}\\
        m_{D_+}^2L_{\mathrm{AdS}}^2&=\tfrac{120}{49}\,,&\qquad\Delta_{D_+}&=\tfrac{20}{7}\qquad&&\text{when }\tau_3=0\,.
    \end{aligned}
    \label{eq:iib-universal}
\end{equation}
Both exact dimensions were previously found in the two-modulus volume–dilaton truncation of~\cite{Apers:2022vfp}. The analysis above shows that $\Delta=4$ persists as an exact eigenvalue of the full scalar Hessian for arbitrary torsion classes, whereas $\Delta=20/7$ does so when $\tau_3=0$. The solvmanifold and nilmanifold vacua of~\cite{VanHemelryck:2025qok} have nonvanishing $\tau_3$, and indeed their spectra contain $\Delta=4$ but not $\Delta=20/7$. 

The value $\Delta=4$ coincides with $\Delta=2(d-1)$ for $d=3$, the dimension of the breathing mode of a generalized Freund--Rubin compactification derived in Appendix~A of~\cite{Apers:2026ahu}. This value also lies at the level-$k=2$ polynomial-shift-symmetric mass of a free scalar in AdS$_3$; its relation to simple large-flux brane realizations was discussed in~\cite{Apers:2022vfp}. The Freund--Rubin result follows from a one-modulus potential, the setting used in the flux-backtracking approach~\cite{Apers:2025pon}; here it follows from an exact eigenmode of the full superpotential Hessian at every supersymmetric type IIB, type I, and heterotic vacuum, at finite flux and without assuming a Freund--Rubin form. 

\subsection{Moduli stabilization and rank constraints}
\label{subsec:iib-axions}

\paragraph{$C_4$ and $B_2$ axions.} The non-closure of the odd three-forms gauges part of the $C_4$ axion sector. The gauge transformations and the covariant derivative inherited from $\delta C_4=\dd(\lambda_\rho\Phi_-^\rho)$ are
\begin{equation}
    \delta A_\rho=\dd\lambda_\rho\,,
    \qquad
    \delta c_4^\sigma=M^{\sigma\rho}\lambda_\rho\,,
    \qquad
    Dc_4^\sigma=\dd c_4^\sigma-M^{\sigma\rho}A_\rho\,,
    \label{eq:iib-c4-gauging}
\end{equation}
and the superpotential is gauge invariant by \eqref{nilpotency} and \eqref{eq:iib-H3-Bianchi-components}. Self-duality of $\widetilde F_5$ identifies the field strength of $A_\rho$ with $Dc_4^\rho$, so the vectors and the axions are the same degrees of freedom, and the $C_4$ sector contributes $b_3^-$ scalars in total rather than $2b_3^-$. Consequently the $\operatorname{rank}M^{\rho\sigma}$ combinations $\delta c_4\in\operatorname{im}M^{\rho\sigma}$ are eaten and the corresponding vectors become massive. The remaining $b_3^--\operatorname{rank}M^{\rho\sigma}$ vectors, stay massless and are dual to the physical $c_4^\rho$.
 
The physical $c_4^\rho$ enter the supersymmetry equations only through $N_{a\rho}c_4^\rho$ in the $B_2$ F-term \eqref{IIBEomNS}; their coupling $c_4^\rho(h_\rho+b^aN_{a\rho})$ in $P_{F_7}$ vanishes at the vacuum by \eqref{IIBEomR4}. Hence $\operatorname{rank}N_{a\rho}$ combinations are stabilized, while the directions in $\ker N_{a\rho}$ do not appear in the supersymmetry equations. From \eqref{nilpotency}, this kernel contains the eaten directions. Therefore, the physical $C_4$ axions are stabilized if and only if
\begin{equation}
    \operatorname{rank}M^{\rho\sigma}+\operatorname{rank}N_{a\rho}=b_3^-\,.
    \label{eq:C4-full-stabilization}
\end{equation}
 
The $B_2$ axions are not gauged, but they are not all fixed by the same mechanism. They enter the superpotential through $b^aN_{a\rho}$, $b^ae_a$ and $\tfrac12\kappa_{abi}b^ab^bf^i$, so their F-term \eqref{IIBEomNS} together with the $C_4$ F-term \eqref{IIBEomR4} forms a system that is linear in $(b^a,c_4^\rho)$ and contains neither the dilaton nor the saxions nor the $C_2$ axions. It can be written as one symmetric linear system with source $(e_a,h_\rho)$, and its solution is
\begin{equation}
    b^a=-(N^{-1})^{\rho a}h_\rho\,,\qquad
    c_4^\rho=(N^{-1})^{\rho a}\left(e_a+\kappa_{abi}f^ib^b\right),\label{C4B2Sol}
\end{equation}
with $(N^{-1})^{\rho a}$ denotes the inverse on the support (the pseudoinverse). Equation \eqref{C4B2Sol} determines only up to $\ker N_{a\rho}$: the remaining $b_2^--\operatorname{rank}N_{a\rho}$ directions, with $\delta b^aN_{a\rho}=0$, are fixed only by the $\kappa_{abi}f^i$ block. Along any direction in the common kernel of $\kappa_{abi}f^i$ and $N_{a\rho}$ the superpotential is invariant at the vacuum, so these directions are perturbatively flat. All $B_2$ axions are therefore stabilized if and only if
\begin{equation}
    \operatorname{rank}\begin{pmatrix}\kappa_{abi}f^i&N_{a\sigma}\\N_{b\rho}&0\end{pmatrix}=b_2^-+\operatorname{rank}N_{a\rho}\,,
    \label{eq:B2-full-stabilization}
\end{equation}
which says that $\kappa_{abi}f^i$ is nondegenerate on the $b_2^--\operatorname{rank}N_{a\rho}$ directions with $\delta b^aN_{a\rho}=0$. 
 
\paragraph{$C_2$ axions.} The $c_2^\alpha$ are not gauged and enter only through the coefficient $f^i+c_2^\alpha N_\alpha{}^i$ of $\widetilde F_3^{\mathrm{int}}$. Writing $\widetilde F_7^{\mathrm{int}}$ for the coefficient in \eqref{iib: F_7}, the saxion equation \eqref{IIBEomMetric} can be put in the form, using the Sherman--Morrison formula and \eqref{IIBEomDilaton},
\begin{equation}
    f^i+c_2^\alpha N_\alpha{}^i
    =\frac{e^{-\phi/2}}{\cV}\left[\frac{s^kM_{k\ell}s^\ell}{6}\,s^i-(G_+^{-1})^{ij}M_{jk}s^k\right].
    \label{eq:IIB-saxion-stabilization}
\end{equation}
Both terms of \eqref{eq:IIB-saxion-stabilization} are built from $s^i$ and $M_{jk}s^k$. Lowering the index with $G^+_{ij}$ and contracting with $N_\beta{}^i$ therefore removes the right-hand side. With $\mathcal N^+_{\alpha\beta}=N_\alpha{}^iG^+_{ij}N_\beta{}^j$, which has rank $\operatorname{rank}N_\alpha{}^i$, the $\operatorname{rank}N_\alpha{}^i$ combinations of the $c_2^\alpha$ are then fixed:
\begin{equation}
    c_2^\alpha=-\left(\mathcal N_+^{-1}\right)^{\alpha\beta}N_\beta{}^iG^+_{ij}f^j\,,
    \label{eq:C2-solution}
\end{equation}
with the inverse on the support of $\mathcal N^+$. The directions with $\delta c_2^\alpha N_\alpha{}^i=0$ do not appear in the superpotential and are perturbatively flat, so all $c_2^\alpha$ are stabilized if and only if $\operatorname{rank}N_\alpha{}^i=b_2^+$. Unlike $b^a$ and $c_4^\rho$, the stabilized combinations depend on the saxions through $G^+_{ij}$.

\paragraph{Saxions and dilaton.} The geometric moduli are stabilized using the $C_2$ F-term \eqref{IIBEomR2} and the remaining equations in \eqref{IIBEomMetric}. The former states $N_\alpha{}^iG^+_{ij}s^j=0$, that is, $N_\alpha{}^i\psi_i=0$, so $\psi_i\in\ker N_\alpha{}^i$. Since $\psi_i$ must be nonzero, a necessary condition for a solution is $\dim\ker N_\alpha{}^i\geq1$, or equivalently,
\begin{equation}
    \operatorname{rank}N_\alpha{}^i<b_3^+\,.
    \label{IIBm<b3}
\end{equation}
Unlike type IIA theory, the saxion cannot be expanded on a basis of the kernel of $N_\alpha{}^i$. Instead, we let $m'=b_3^+-\operatorname{rank}N_\alpha{}^i$ and let $u_{ri}$, $r=1,\ldots,m'$, be a fixed basis of this kernel, $N_\alpha{}^iu_{ri}=0$. Since $s^i\mapsto\psi_i$ is invertible~\cite{HitchinStableForms}, we parametrize the geometric moduli through $\psi_i=t^ru_{ri}$, where the $m'$ real coefficients $t^r$ are to be determined; $s^i$, $G^+_{ij}$ and $\cV$ are then functions of the $t^r$. Substituting the closed form \eqref{eq:C2-solution} for $c_2^\alpha$ into \eqref{IIBEomMetric} turns the dressed flux into the projection of $f^i$ orthogonal, in the metric $G^+$, to the vectors $N_\alpha{}^i$,
\begin{equation}
    f^i+c_2^\alpha N_\alpha{}^i=f^i-N_\alpha{}^i(\mathcal N_+^{-1})^{\alpha\beta}N_\beta{}^jG^+_{jk}f^k\,.
\end{equation}
Contracting equation \eqref{IIBEomMetric} with the vectors $(G_+^{-1})^{ij}u_{rj}$, the projector term drops because $u_{rj}N_\alpha{}^j=0$, and the equation is written as
\begin{align}
    0={}&\left[e^{\phi/2}\psi_if^i-2e^{-\phi/2}\left(f_7+b^ae_a+\tfrac12\kappa_{abi}f^ib^ab^b\right)-s^kM_{k\ell}s^\ell\right]u_{ri}s^i\nonumber\\
    &+3e^{\phi/2}\cV\,u_{ri}f^i+3\,u_{ri}(G_+^{-1})^{ij}M_{jk}s^k\,,
    \label{eq:iib-kernel-saxion}
\end{align}
with $b^a$ given by \eqref{C4B2Sol}; the $c_4^\rho$ term of the seven-form coefficient has dropped by the $C_4$ F-term. The $c_2^\alpha$ dependence has dropped out. These $m'$ equations generically fix the $m'$ coefficients $t^r$, and hence the geometric moduli $s^i$. Given the geometric moduli, one can further fix the dilaton using \eqref{IIBEomDilaton}. 

\subsection{\texorpdfstring{Singlet-torsion $B_2$-axion dimensions}{Singlet-torsion B2-axion dimensions}}
\label{subsec:iib-singlet-torsion-b-spectrum}

The $B_2$ axion has a universal dimension in type IIB theory as well, albeit on a restricted supersymmetric branch. We assume that the torsion matrices connecting the orientifold-odd forms vanish, $N_{a\rho}=M^{\rho\sigma}=0$, and that $\tau_3=0$. Since $\tau_1=\tau_2=0$ on-shell, the condition $\tau_3=0$ is equivalent to the intrinsic torsion being purely singlet. We take $\tau_{0,\star}\neq0$ below, so that the vacuum is AdS$_3$.

The condition $\tau_3=0$ implies $\dd\Phi=\tau_0\Psi$. Using \eqref{torsion-matrices} and \eqref{g2-structure-IIB} we obtain
\begin{equation}
    M_{ij}s^j=\tau_0\cV G^+_{ij}s^j\,,\qquad s^iM_{ij}s^j=7\tau_0\cV\,.
    \label{eq:iib-singlet-torsion-relations}
\end{equation}
Thus, on this branch the geometric contribution is entirely determined by the singlet torsion. Together with the flux balance \eqref{eq:iib-universal-flux-balance}, the on-shell superpotential is
\begin{equation}
    P_\star=\frac{1}{3}P_{\mathrm{geom},\star}=\frac{1}{24\cV_\star^2}s_\star^iM_{ij}s_\star^j=\frac{7\tau_{0,\star}}{24\cV_\star}\,.
    \label{eq:iib-singlet-torsion-Pstar}
\end{equation}
Using \eqref{eq:iib-universal-flux-balance}, the saxion F-term \eqref{IIBEomMetric} can be written at this vacuum as
\begin{align}
    0={}&M_{ij}s_\star^j+e^{\phi_\star/2}\cV_\star G^+_{ij,\star}\left(f^j+c_{2,\star}^\alpha N_\alpha{}^j\right)-\frac{1}{6}G^+_{ij,\star}s_\star^j\left(s_\star^kM_{k\ell}s_\star^\ell\right) \\
    ={}&e^{\phi_\star/2}\cV_\star G^+_{ij,\star}\left(f^j+c_{2,\star}^\alpha N_\alpha{}^j\right)-\frac{1}{6}\tau_{0,\star}\cV_\star G^+_{ij,\star}s_\star^j\,.
\end{align}
Since $G^+_{ij,\star}$ is nondegenerate, this gives
\begin{equation}
    f^i+c_{2,\star}^\alpha N_\alpha{}^i=\frac{\tau_{0,\star}}{6}e^{-\phi_\star/2}s_\star^i\,.
    \label{eq:iib-singlet-torsion-F3-alignment}
\end{equation}
The axion-dressed three-form flux vector therefore aligns with the saxions $s^i$ at the vacuum.

We now determine the $B_2$-axion masses. For $N_{a\rho}=0$, the $B_2$ F-term \eqref{IIBEomNS} depends on the other scalar fields only through its overall prefactor. Since its bracket vanishes at the supersymmetric vacuum, all mixed superpotential-Hessian components involving $b^a$ vanish. The nonzero $B_2$ block is
\begin{equation}
    \left.\nabla_{b^a}\partial_{b^b}P_{\mathrm{IIB}}\right|_\star=\frac{e^{-\phi_\star/2}}{4\cV_\star^2}\kappa_{abi}f^i\,.
    \label{eq:iib-singlet-torsion-b-hessian-first}
\end{equation}
The Stokes relation \eqref{stokes}, together with $N_{a\rho}=0$, implies $\kappa_{abi}N_\alpha{}^i=0$. We can therefore use \eqref{eq:iib-singlet-torsion-F3-alignment} and \eqref{eq:metric-intersection-identity} to obtain
\begin{align}
    \left.\nabla_{b^a}\partial_{b^b}P_{\mathrm{IIB}}\right|_\star&=\frac{e^{-\phi_\star/2}}{4\cV_\star^2}\kappa_{abi}\left(f^i+c_{2,\star}^\alpha N_\alpha{}^i\right) =\frac{\tau_{0,\star}e^{-\phi_\star}}{24\cV_\star^2}\kappa_{abi}s_\star^i =-\frac{\tau_{0,\star}e^{-\phi_\star}}{24\cV_\star}G^-_{ab,\star}\,.
    \label{eq:iib-singlet-torsion-b-hessian}
\end{align}
Thus, the $B_2$ superpotential Hessian is proportional to the $B_2$ kinetic metric $\mathcal G^{(b)}_{ab}=\frac14e^{-\phi}G^-_{ab}$. Raising one index and using \eqref{eq:iib-singlet-torsion-Pstar} gives
\begin{align}
    \mathcal H^a{}_b&=\left.4e^\phi(G_-^{-1})^{ac}\nabla_{b^c}\partial_{b^b}P_{\mathrm{IIB}}\right|_\star=-\frac{\tau_{0,\star}}{6\cV_\star}\delta^a{}_b\,,\qquad\frac{\lambda_b}{2P_\star}=-\frac{2}{7}\,,
\end{align}
and equations \eqref{eq:general-normalized-Hessian-mass} and \eqref{eq:general-conformal-dimension} then give
\begin{equation}
    m_b^2L_{\mathrm{AdS}}^2=\frac{32}{49}\,,\qquad \Delta_b=\frac{16}{7}\,.
    \label{eq:iib-flux-aligned-b-spectrum}
\end{equation}
Thus, all $b_2^-$ modes in this $B_2$ sector have the same exact noninteger conformal dimension.

There is, however, a caveat. The $b^a$ also mix kinetically with the $c_4^\rho$ when $c_{2,\star}^\alpha\neq0$, so we must check that this does not change \eqref{eq:iib-flux-aligned-b-spectrum}. Self-duality of $\widetilde F_5$ and the Chern--Simons term in its Bianchi identity give the one-external-leg component $\widetilde F_5\supset(Dc_4^\rho-c_2^\alpha\kappa_{a\alpha}{}^{\rho}\dd b^a)\wedge\Psi^-_\rho$~\cite{Aikot:2026tzn}, so that, with $M^{\rho\sigma}=0$ on this branch,
\begin{equation}
    \begin{split}
        (4\pi\sqrt{-g_3})^{-1}\cL_{b,c_4}
        &=-\tfrac14e^{-\phi}G^-_{ab}\partial_\mu b^a\partial^\mu b^b\\
        &-\tfrac14\widehat G^-_{\rho\sigma}\!\left(\partial_\mu c_4^\rho-c_2^\alpha\kappa_{a\alpha}{}^{\rho}\partial_\mu b^a\right)\!\left(\partial^\mu c_4^\sigma-c_2^\beta\kappa_{b\beta}{}^{\sigma}\partial^\mu b^b\right).
        \end{split}
    \label{eq:iib-b-c4-kinetic}
\end{equation}
Since $M^{\rho\sigma}=0$, the $c_4^\rho$ are not St\"uckelberg fields on this branch, and the $C_4$ F-term together with $N_{a\rho}=0$ gives $h_\rho=0$, so they do not appear in the scalar potential. For every nonzero-mass eigenmode, the $c_4^\rho$ component of the generalized eigenvalue equation then requires
\begin{equation}
    \delta c_4^\rho=c_{2,\star}^\alpha\kappa_{a\alpha}{}^\rho\,\delta b^a\,.
    \label{eq:iib-singlet-torsion-b-c4-eigenvector}
\end{equation}
Along these modes the second square in \eqref{eq:iib-b-c4-kinetic} vanishes, so the induced kinetic metric is again $\mathcal G^{(b)}_{ab,\star}$ and the eigenvalue is unchanged. The mixing therefore changes the eigenvectors but not the eigenvalues. For $c_{2,\star}^\alpha=0$ the eigenvectors are purely $B_2$ fluctuations, while for $c_{2,\star}^\alpha\neq0$ they contain the $C_4$ components \eqref{eq:iib-singlet-torsion-b-c4-eigenvector}. The remaining $C_4$ combinations do not enter the potential and are perturbatively flat, with $\Delta=2$.

\subsection{The type I truncation and D9-brane gauge bundles}
\label{subsec:type-I-truncation}

The O9 projection removes the $\Omega_p$-odd fields:
\begin{equation}
    B_2
    =
    0\,,
    \qquad
    C_0
    =
    0\,,
    \qquad
    C_4
    =
    0\,,
    \qquad
    \widetilde F_5
    =
    0\,.
    \label{eq:type-I-projection}
\end{equation}
Consequently, the strict closed-string type I superpotential is 
\begin{equation}
    P_{\mathrm{I}}
    =
    \frac{e^{-\phi/2}}{4\cV^2}f_7
    -
    \frac{e^{\phi/2}}{4\cV^2}
    \psi_i(s)
    \left(f^i+c_2^\alpha N_\alpha{}^i\right)
    +
    \frac{1}{8\cV^2}s^iM_{ij}s^j\,.
    \label{eq:type-I-component-superpotential}
\end{equation}
The universal relation \eqref{eq:iib-universal-flux-balance} and the conclusion that singlet torsion is necessary for a supersymmetric AdS$_3$ vacuum continue to hold without modification.

The O9-plane charge and tension are canceled locally by 32 D9-branes~\cite{GreenSchwarzWittenBook,PolchinskiVol2}. Their gauge bundle does not generate a separate additional term in \eqref{eq:type-I-component-superpotential}. Instead, the type I RR three-form is replaced by its Chern--Simons-completed expression
\begin{equation}
    \widetilde F_3^{\mathrm{loc}}
    =
    \widetilde F_3
    +
    \frac{\alpha'}{4}
    \left(
        \omega_{3\mathrm{YM}}(A)
        -
        \omega_{3L}(\Omega_+)
    \right)\,.
    \label{eq:type-I-CS-completed-F3}
\end{equation}
Here the Yang-Mills and Lorentz Chern-Simons three-forms $\omega_\text{3YM}$ and $\omega_\text{3L}$, built from the gauge connection $A$ and torsionful spin connection $\Omega_+$, follow the conventions of~\cite{Aikot:2026tzn}. Its Bianchi identity is
\begin{equation}
    \dd\widetilde F_3^{\mathrm{loc}}
    =
    \frac{\alpha'}{4}
    \left[
        \Tr\left(F_2\wedge F_2\right)
        -
        \Tr\left(R_{2,+}\wedge R_{2,+}\right)
    \right]
    +
    j_{\mathrm{O5/D5}}\,.
    \label{eq:type-I-CS-Bianchi}
\end{equation}
Thus, gauge and gravitational instanton numbers modify the tadpole \eqref{eq:iib-O5-tadpole-components} and the effective periods entering the $F_3$ sector. A complete type I vacuum must additionally solve the bundle supersymmetry conditions and include any bundle moduli in the physical mass matrix; the closed-sector results above do not by themselves establish stabilization of these open-string fields. 

By S-duality, the type I supersymmetry equations and their solutions are very similar to the results for the heterotic string theory. We direct the reader to the next section~\ref{sec:heterotic} for a discussion of moduli stabilization and universal results. 

\subsection{Exact flat direction in supersymmetric Minkowski vacua}
\label{subsec-iib-minkowski}

As in section~\ref{subsec-iia-minkowski}, we now show that every finite supersymmetric Minkowski vacuum retains an exact classical volume--dilaton flat direction, including intrinsic torsion and the retained axionic sector.

At a supersymmetric Minkowski vacuum, $V_\star=P_{\mathrm{IIB},\star}=0$, and the universal relation \eqref{eq:iib-universal-flux-balance} requires
\begin{equation}
    P_{F_7,\star}=P_{F_3,\star}=P_{\mathrm{geom},\star}=0\,.
\end{equation}
The first condition sets the axion-dressed seven-form coefficient in \eqref{eq:iib-component-superpotential} to zero. Since $b^a$ and $c_4^\rho$ enter only $P_{F_7}$, their supersymmetry equations also set its derivatives along all physical axionic directions to zero; $P_{F_7}$ is independent of $c_2^\alpha$. Its dilaton and saxion derivatives vanish because they are proportional to $P_{F_7}$ itself. Thus, $P_{F_7}$ and all its first derivatives vanish separately at the vacuum.

Combining the dilaton and volume weights displayed in \eqref{eq:iib-dilaton-balance} and \eqref{eq:iib-volume-balance}, we consider the rescaling
\begin{equation}
    s^i\longrightarrow e^{3t/7}s^i\,,\qquad \cV\longrightarrow e^t\cV\,,\qquad \phi\longrightarrow\phi+\frac47t\,,
\end{equation}
with all axions, shape ratios, fluxes, and torsion data fixed. Under this transformation, $P_{F_7}$ acquires the factor $e^{-16t/7}$, while $P_{F_3}$ and $P_{\mathrm{geom}}$ both acquire $e^{-8t/7}$. The vanishing seven-form coefficient and its axion derivatives remain unchanged, so $P_{F_7}$ and its first derivatives stay zero along the trajectory. The two remaining terms scale identically, and hence $P_{\mathrm{IIB}}=0$ and all supersymmetry equations $\partial_M P_{\mathrm{IIB}}=0$ remain satisfied. Thus every finite supersymmetric Minkowski solution belongs to a continuous one-parameter family, and one combination of the overall volume and dilaton is necessarily an exact classical flat direction. The same conclusion applies to the strict closed-string type I truncation of section~\ref{subsec:type-I-truncation}.

\section{Heterotic string theory}
\label{sec:heterotic}

In this section, we study supersymmetric heterotic compactifications on $G_2$-structure spaces, focusing on the universal closed-string sector, common to both the $\mathrm{SO}(32)$ and $E_8\times E_8$ theories. We analyze the axion stabilization and rank constraints and show that the universal vacuum relations persist from the type IIB/type I analysis. We also discuss the bundle conditions and Green--Schwarz gaugings and briefly recall the known anisotropic family.

\subsection{Gauge-invariant flux and the heterotic Bianchi identity}
\label{subsec:heterotic-setup}

In the heterotic theory, no orientifold projection is required and the gauge group enters through the choice of bundle, the normalization of the trace, and the bundle-dependent scalar and vector sectors. We thus choose the basis, torsion matrices, triple-intersection numbers, and the $G_2$ structure as in section~\ref{subsec:g2-notation}.

To first order in $\alpha'$, the terms that distinguish the heterotic Einstein-frame action from the universal NSNS action are given by~\cite{GreenSchwarzWittenBook,PolchinskiVol2,BergshoeffDeRoo1989}
\begin{equation}
    S_{\mathrm{het}}
    \supset
    -\frac{1}{2\kappa_{10}^2}
    \int\dd^{10}x\sqrt{-g_{10}}\,
    \frac{\alpha'}{4}e^{-\phi/2}
    \left(
        \Tr|F_2|^2
        -
        \Tr|R_{2,+}|^2
    \right),
    \label{eq:het-alpha-prime-action}
\end{equation}
where $R_{2,+}$ is the associated curvature of $\Omega_+=\omega+\frac 12H_3$, the torsionful spin connection. A different allowed choice of tangent-bundle connection amounts to a field redefinition at this order and changes the effective action only at the next order in the $\alpha'$ expansion~\cite{BergshoeffDeRoo1989,delaOssa:2017pqy}.

The field strength that enters the action and the supersymmetry equations is the gauge-invariant three-form
\begin{equation}
     H_3^\text{loc}
    =
    \dd B_2
    +
    \frac{\alpha'}{4}
    \left[
        \omega_{3\mathrm{YM}}(A)
        -
        \omega_{3L}(\Omega_+)
    \right],
    \label{eq:het-gauge-invariant-H}
\end{equation}
where $A$ is the gauge connection. The possible globally nontrivial flux data are understood as part of the differential-cohomology class of $H_3$.

We choose the unprojected truncated bases $\{\Upsilon_A,\Phi_I,\Psi^I,T^A\}$ satisfying the equations \eqref{orthonormality}--\eqref{eq:metric-intersection-identity} above. The $G_2$ structure and the ten-dimensional potentials are
\begin{equation}
    \Phi=s^I\Phi_I\,,\qquad B_2=b^A\Upsilon_A+B_2^{(3)}\,.
    \label{eq:het-B2-expansion}
\end{equation}
Here, $B_2^{(3)}$ is external and understood to be nondynamical and dual to an internal 7-form flux.

At fixed gauge and tangent-bundle connections, the internal field strengths may be expanded as
\begin{align}
    H^\text{int}_3&=(h^I+b^AN_A{}^I)\Phi_I\,,\label{eq:het-H3-internal}\\
    H^\text{int}_7&=h_7\,\widetilde{\text{dvol}}_7\,.\label{eq:het-H7-internal}
\end{align}
The coefficients $h^I$ include the connection-dependent contribution needed to describe the chosen gauge-invariant representative. They are therefore not, in general, independent fluxes. Their quantization is the shifted heterotic flux quantization determined by the gauge bundle, the tangent bundle, and the choice of trivialization of the Bianchi identity.

The anomaly-canceling Bianchi identity takes the standard form~\cite{GreenSchwarzWittenBook,PolchinskiVol2}
\begin{equation}
    \dd H_3^\text{loc}
    =
    \frac{\alpha'}{4}
    \left[
        \Tr(F_2\wedge F_2)
        -
        \Tr(R_{2,+}\wedge R_{2,+})
    \right]
    +
    j_\text{NS5}\,.
    \label{eq:het-Bianchi-identity}
\end{equation}
Upon integration, we find the tadpole condition
\begin{equation}
    h^IM_{IJ}=\frac{\alpha'}4\int_X\Phi_J\wedge\left(\Tr(F_2\wedge F_2)-\Tr(R_{2,+}\wedge R_{2,+})\right)+q_{5,J}\,.
    \label{eq:het-Bianchi-components}
\end{equation}

The heterotic real superpotential in the present Einstein-frame normalization is
\begin{equation}
    P_{\mathrm{het}}
    =
    \frac{1}{4\cV^2}
    \int_X
    \left(
        e^{\phi/2}H_7^{\mathrm{int}}
        -
        e^{-\phi/2}\Psi\wedge H_3^{\mathrm{loc, int}}
        +
        \frac{1}{2}\Phi\wedge\dd_7\Phi
    \right)
    +
    \mathcal O(\alpha'^2).
    \label{eq:het-form-superpotential}
\end{equation}
This is the Einstein-frame form of the heterotic $G_2$ superpotential derived in~\cite{delaOssa:2019cci} and used in~\cite{Tringas:2025bwe,Aikot:2026tzn}. The first and second variations of this superpotential, their relation to the supersymmetry conditions and infinitesimal moduli, and the associated one-loop partition function were studied in~\cite{delaOssa:2024lwx}. For the $\mathrm{SO}(32)$ theory the above is also the S-dual of the type I superpotential \eqref{eq:type-I-component-superpotential}, under $\phi_{\mathrm{het}}=-\phi_{\mathrm{I}}$ and the exchange of the type I RR fluxes with the heterotic NS-NS fluxes~\cite{Witten:1995ex,Polchinski:1995df,GreenSchwarzWittenBook,PolchinskiVol2}. This S-duality check does not apply to the $E_8\times E_8$ theory, although its universal superpotential and closed-string scalar metric have the same form. The corresponding type I results are discussed in sections~\ref{subsec:iib-flux-balance}, \ref{subsec:iib-axions}, \ref{subsec:type-I-truncation}, and~\ref{subsec-iib-minkowski}; the heterotic consequences are collected below.

At fixed bundle data, the component form is
\begin{equation}
    P_{\mathrm{het}}
    =
    \underbrace{
        \frac{e^{\phi/2}}{4\cV^2}h_7
    }_{P_{H_7}}
    +
    \underbrace{
        -
        \frac{e^{-\phi/2}}{4\cV^2}
        \psi_I(s)
        \left(
            h^I+b^A N_A{}^I
        \right)
    }_{P_{H_3}}
    +
    \underbrace{
        \frac{1}{8\cV^2}
        s^I M_{IJ}s^J
    }_{P_{\mathrm{geom}}}
    +
    \mathcal O(\alpha'^2).
    \label{eq:het-component-superpotential}
\end{equation}
The connection dependence of $P_{H_3}$ is implicit in \eqref{eq:het-component-superpotential} and is kept explicit in the form expression \eqref{eq:het-form-superpotential}. The gauge bundle does not produce an additional independent term proportional to $\Phi\wedge\Tr(F_2\wedge F_2)$; its effects enter through the complete three-form $H^\text{loc}_3$, the Bianchi identity, the scalar metric, and the bundle-field variations below.

The leading-order supersymmetric equations of motion require the following derivatives to vanish
\begin{align}
    \partial_\phi P_\text{het}&=\frac{e^{\phi/2}}{8\mathcal V^2}h_7+\frac{e^{-\phi/2}}{8\mathcal V}G_{IJ}h^Is^J\,,\label{HetEomDilaton}\\
    \begin{split}
            \partial_IP_\text{het}&=-\frac{e^{\phi/2}}{6\mathcal V^2}G_{IJ}s^Jh_7+\frac{e^{-\phi/2}}{4\mathcal V}\left(\frac 13G_{IJ}G_{KL}s^Js^K+G_{IL}\right)(N_A{}^Lb^A+h^L)\\
            &\qquad+\frac 1{4\mathcal V^2}\left(M_{IJ}s^J-\frac 13G_{IJ}M_{KL}s^Js^Ks^L\right)\,,
    \end{split}\label{HetEomMetric}\\
    \partial_AP_\text{het}&=-\frac{e^{-\phi/2}}{4\mathcal V}G_{IJ}N_A{}^Is^J\label{HetEomNS}\,,
\end{align}
where we used $\partial_AP_\text{het}=0$ to simplify $\partial_\phi P_\text{het}$.

A complete supersymmetric solution must also satisfy the $G_2$-instanton conditions, obtained by varying \eqref{eq:het-form-superpotential} with respect to the
gauge and tangent-bundle connections~\cite{delaOssa:2017pqy,delaOssa:2019cci,Aikot:2026tzn}:
\begin{equation}
    \Psi\wedge F_2=0\,,
    \qquad
    \Psi\wedge R_{2,+}=0\,.
    \label{eq:het-G2-instanton-conditions}
\end{equation}
Since $\Psi$ depends
on the geometric moduli, the instanton conditions also constrain
the geometry even when the bundle connections are held fixed.
The physical deformation problem, including the removal of
spurious tangent-bundle connection deformations, was formulated
in~\cite{McOrist:2024ivz}.

\subsection{Moduli stabilization and rank constraints}
\label{subsec:het-axions}

At fixed bundle data, the heterotic component superpotential \eqref{eq:het-component-superpotential} is the type I superpotential \eqref{eq:type-I-component-superpotential} under $\phi\to-\phi$ and $(f_7,f^i,c_2^\alpha)\to(h_7,h^I,b^A)$, with the parity restrictions removed. The stabilization problem is therefore that of subsection~\ref{subsec:iib-axions} with the orientifold-odd sector absent, that is, $b^a=c_4^\rho=e_a=0$ and $\widetilde F_7^{\mathrm{int}}\to h_7$. We quote the results.

\paragraph{$B_2$ axions.} The dilaton and saxion equations \eqref{HetEomDilaton} and \eqref{HetEomMetric} determine the dressed flux, the counterpart of \eqref{eq:IIB-saxion-stabilization},
\begin{equation}
    h^I+b^AN_A{}^I=\frac{e^{\phi/2}}{\cV}\left[\frac{s^KM_{KL}s^L}{6}\,s^I-(G^{-1})^{IJ}M_{JK}s^K\right].
    \label{eq:het-hhat-solution}
\end{equation}
Lowering the index with $G_{IJ}$ and contracting with $N_A{}^I$ removes the right-hand side, by \eqref{HetEomNS} and $N_A{}^IM_{IJ}=0$, so
\begin{equation}
    b^A=-\left(\mathcal N^{-1}\right)^{AB}N_B{}^IG_{IJ}h^J\,,\qquad
    \mathcal N_{AB}=N_A{}^IG_{IJ}N_B{}^J\,.
    \label{eq:het-b-solution}
\end{equation}
$\mathcal N_{AB}$ is a Gram matrix in the positive-definite metric $G_{IJ}$, so all $B_2$ axions are stabilized if and only if
\begin{equation}
    \operatorname{rank}N_A{}^I=b_2\,,
    \label{eq:het-b-rank-condition}
\end{equation}
where $b_2$ and $b_3$ count the retained two- and three-forms. Otherwise $b_2-\operatorname{rank}N_A{}^I$ axions stay perturbatively flat.

Bundle gauge fields can still gauge the $B_2$ axions. For a retained abelian vector $A^\Lambda$ with an unbroken generator $T_\Lambda$, the Yang--Mills Chern--Simons term in \eqref{eq:het-gauge-invariant-H} gives
\begin{equation}
    Db^A=\dd b^A+k_\Lambda{}^AA^\Lambda\,,\qquad
    \frac{\alpha'}{2}\Tr\left(T_\Lambda F_2^{\mathrm{bg}}\right)=k_\Lambda{}^A\Upsilon_A\,.
    \label{eq:het-GS-charge-and-gauging}
\end{equation}
The projected flux is closed, so $k_\Lambda{}^AN_A{}^I=0$, and the gauged directions lie in the shift kernel. The perturbatively flat axions then number
\begin{equation}
    n_{b,\mathrm{flat}}=b_2-\operatorname{rank}N_A{}^I-\operatorname{rank}k\,,
    \label{eq:het-physical-flat-b-axions}
\end{equation}
the removed directions being St\"uckelberg fields rather than moduli. The rank condition above applies when $k=0$.

\paragraph{Saxions and dilaton.} The $B_2$ F-term \eqref{HetEomNS} states $N_A{}^I\psi_I=0$, so a nonzero solution requires 
\begin{equation}
    \operatorname{rank}N_A{}^I=b_2<b_3\label{eq:het-saxion-rank-condition}
\end{equation}
The same constraint is imposed for AdS$_3$ vacua. An AdS$_3$ vacuum requires $\text{rank }M_{IJ}\ge 1$, and nilpotency requires $\text{rank }N+\text{rank }M\le b_3$. Substituting \eqref{eq:het-b-rank-condition}, we retrieve \eqref{eq:het-saxion-rank-condition}.

Let $m'=b_3-\operatorname{rank}N_A{}^I$, let $u_{rI}$, $r=1,\ldots,m'$, be a fixed basis of $\ker N_A{}^I$, and parametrize $\psi_I=t^ru_{rI}$ with $t^r$ to be determined. Substituting \eqref{eq:het-b-solution} into \eqref{HetEomMetric} and contracting with $(G^{-1})^{IJ}u_{rJ}$, the $b^A$ dependence drops, and the equation reads, exactly as in \eqref{eq:iib-kernel-saxion} with $\phi\to-\phi$ and $f_7\to h_7$,
\begin{align}
    0={}&\left[e^{-\phi/2}\psi_Ih^I-2e^{\phi/2}h_7-s^KM_{KL}s^L\right]u_{rI}s^I\nonumber\\
    &+3e^{-\phi/2}\cV\,u_{rI}h^I+3\,u_{rI}(G^{-1})^{IJ}M_{JK}s^K\,.
    \label{eq:het-kernel-saxion}
\end{align}
These $m'$ equations fix the $t^r$, and the dilaton follows from \eqref{HetEomDilaton}.

\subsection{Universal flux balance and flat direction in Minkowski vacua}
\label{subsec:heterotic-flux-balance}

The relations of subsection~\ref{subsec:iib-flux-balance} hold with $F\to H$ and $\phi\to-\phi$, for both gauge groups since they use only the common universal superpotential:
\begin{equation}
    P_{H_3,\star}=P_{H_7,\star}\,,\qquad
    P_{\mathrm{geom},\star}=-3P_{H_7,\star}\,,\qquad
    P_\star=\tfrac13P_{\mathrm{geom},\star}\,,
    \label{eq:het-universal-flux-balance}
\end{equation}
so a supersymmetric AdS$_3$ vacuum requires nonzero integrated singlet torsion. The $B_2$ axions enter only $P_{H_3}$, as $c_2^\alpha$ enter only $P_{F_3}$, so the exactness argument of subsection~\ref{subsec:iib-flux-balance} goes through with $b^a=c_4^\rho=0$, and the volume--dilaton directions $D_{\mathrm{het}}=-4\partial_\phi+3s^I\partial_I$ and $D'_{\mathrm{het}}=28\partial_\phi+3s^I\partial_I$ have
\begin{equation}
    \Delta_{D}=4\quad\text{at every supersymmetric vacuum,}\qquad
    \Delta_{D'}=\tfrac{20}{7}\quad\text{when }\tau_3=0\,.
    \label{eq:het-universal-delta4}
\end{equation}
Likewise a supersymmetric Minkowski vacuum requires $h_7=0$ and carries the flat direction of subsection~\ref{subsec-iib-minkowski},
\begin{equation}
    s^I(t) \longrightarrow e^{3t/7}s^I\,,\qquad \cV(t) \longrightarrow e^t\cV\,,\qquad \phi(t) \longrightarrow \phi-\tfrac47t\,.
    \label{eq:het-Minkowski-flat-direction}
\end{equation}
For fixed nonzero torsion and a fixed nonsingular shape, $|V_\star|$ and $m_{\mathrm{KK}}^2$ both scale as $\cV^{-16/7}$, so the isotropic obstruction to scale separation of subsection~\ref{subsec:iib-flux-balance} applies. The anisotropic family of~\cite{Tringas:2025bwe} evades it: its radius estimates give $L_{\mathrm{KK}}/L_{\mathrm{AdS}}\sim N^{-1}$ for $1\ll N^5\ll h_7$, and its eight retained scalars have seven dimensions $\Delta=4$ and one $\Delta=12$. These results use a left-invariant truncation and smeared gravitational instantons, and do not establish stabilization of the omitted twisted and bundle modes.

\section{Conclusions and outlook}
\label{sec:conclusions}

Using the general framework for $G_2$ compactifications developed in \cite{Aikot:2026tzn}, we analyze supersymmetric solutions of the resulting three-dimensional theories for type IIA, type IIB/type I, and heterotic string theory, without specifying a particular internal manifold. We summarize the main results below, grouping them by the mechanisms underlying each solution.

\paragraph{Vacuum structure and stabilization.} The dilaton and volume equations fix the ratios of the superpotential sectors in each theory: $3:-4:2$ for $(H,F_4,F_0)$ in massive type IIA and $1:1:-3$ for $(F_7,F_3,\text{geometric})$ in type IIB/type I, $(H_3,H_7,\text{geometric})$ in the heterotic theory. A supersymmetric AdS$_3$ vacuum therefore requires a nonzero Romans mass in type IIA and nonzero singlet torsion in the other theories, while every supersymmetric Minkowski vacuum in all three carries an exact volume--dilaton flat direction, so the retained classical closed-string sector cannot fully stabilize the moduli there.The non-closure of the retained internal forms is encoded in torsion matrices \eqref{torsion-matrices}. The supersymmetry conditions for stabilizing the geometric moduli impose rank constraints on these matrices. Some axionic shifts are gauged by three-dimensional vector fields, so the corresponding axions are not independent physical fields. These gauge redundancies must be taken into account before solving the supersymmetry equations for the physical axions.

\paragraph{Universal conformal dimensions.}
In every supersymmetric AdS$_3$ vacuum of the type IIB/type I and heterotic compactifications considered here on $G_2$-structure spaces, we find a scalar fluctuation involving the overall internal volume and the dilaton whose dual operator has conformal dimension $\Delta=4$, independent of the choice of internal space. The result follows from the superpotential's scaling properties together with the universal scalar metric and the supersymmetry conditions. We also find another scalar direction with $\Delta=20/7$ universally, when $\tau_3=0$, where $\tau_i$ denote the four irreducible components of the $G_2$ intrinsic torsion \eqref{taus}. In type IIA the volume--dilaton plane is not invariant in general, so we obtain only the bound $\Delta_{\max}\geq(20+4\sqrt{67})/7\simeq7.53$. The $B_2$-axion modes have universal conformal dimensions in two cases: when the retained orientifold-odd two-forms are closed in type IIA, $\Delta_b=6$, and on the singlet-torsion branch in type IIB, $\Delta_b=16/7$. The values $4$ and $6$ lie on the polynomial-shift list while $20/7$ and $16/7$ do not, so integrality is not a consequence of minimal supersymmetry; it reflects extra structure of the examples. The exact $\Delta_b=6$ also matters holographically. Together with the perturbatively flat $\Delta=2$ axions it forms a super-extremal $(6;2,2)$ arrangement, for which the criterion of~\cite{Bobev:2025yxp} demands a vanishing cubic coupling, and here this holds because the two-derivative Lagrangian is invariant under reflecting the $B_2$ axions about their vacuum values.

\paragraph{Large flux and scale separation.} The physical large-flux directions are fewer than the naive kernel count, since exact flux components are absorbed by axion shifts, and at nonzero Romans mass a growing $F_2$ flux supplies no independent limit. In type IIA, admissible large-$F_4$ families with fixed sources give weak coupling, large string-frame volume, and $L_{\mathrm{KK}}^2/L_{\mathrm{AdS}}^2\sim n^{-1}$, but a fixed $\tau_2$ ruins this hierarchy, so any surviving $\tau_2$ must be suppressed in the limit. In type IIB/type I and heterotic vacua the obstruction is structural: the singlet torsion required for AdS$_3$ also sets the internal curvature, so an isotropic limit gives an order-one ratio $L_{\mathrm{KK}}^2/L_{\mathrm{AdS}}^2$ independent of the fluxes. Scale separation there needs anisotropy, as in the known families~\cite{VanHemelryck:2025qok,Miao:2025rgf,Tringas:2025bwe}, whose Kaluza--Klein spectrum must be computed from the individual radii. Without torsion, the unwarped type IIB/type I reduction with calibrated sources has only a fluxless Minkowski locus.

\paragraph{Outlook.} The conditional branches, namely massive type IIA with surviving two-form axions and type IIB on the singlet-torsion branch, still lack compact realizations, and explicit examples with nontrivial two- and five-form sectors would test the rank conditions and flux identifications against real geometry. Localized sources, resolution, open-string and bundle moduli~\cite{Junghans:2020acz,Marchesano:2020qvg,Bardzell:2024anh,Emelin:2024vug,Arboleya:2025lwu,delaOssa:2017pqy} are needed to assess stability beyond the retained sectors. Finally, the perturbatively flat directions and the exact dimensions found here must be re-examined once higher-derivative and nonperturbative corrections are included, in particular in the type I and heterotic theories, where the Bianchi identity ties the torsion to gauge and gravitational instanton numbers.

\subsection*{Acknowledgments}
We would like to thank Fotis Farakos, Zheng Miao, Muthusamy Rajaguru, Vincent Van Hemelryck and Thomas Van Riet for discussions and George Tringas for collaboration during the early stages of this project. This work is supported in part by NSF grant PHY-2609419.

\paragraph*{AI assistance.}
The authors used ChatGPT-5.6 Sol, ChatGPT-6 Astra and Claude Opus 5.5 to assist with writing, editing, and checking calculations in this paper. The authors have carefully reviewed all such contributions and take full responsibility for the content, accuracy, and conclusions of the work.

\bibliographystyle{JHEP}
\bibliography{refs}

\end{document}